\documentclass{aa}  

\usepackage{graphicx}
\usepackage{txfonts}
\usepackage{lscape}
\usepackage{pdfpages}
\usepackage{afterpage}
\usepackage{hyperref}
\usepackage{placeins}
\begin{document}

   \title{Stellar angular momentum of intermediate-redshift galaxies in MUSE surveys}

   \author{Constanza Mu\~{n}oz L\'{o}pez
          \inst{1}\fnmsep\thanks{e-mail: \texttt{cmunoz@aip.de}}
          \and
          D. Krajnovi\'{c}
          \inst{1}
          \and
          B. Epinat
          \inst{2,3}
          \and
          Y. Herrero-Alonso
          \inst{1}
          \and
          T. Urrutia
          \inst{1}
          \and
          W. Mercier
          \inst{3,4}                    
          \and
          N. F. Bouch\'{e}
          \inst{5}
          \and
          L. A. Boogaard
          \inst{6}
          \and
          T. Contini
          \inst{4}
          \and
          L. Michel-Dansac
          \inst{3,5}
          \and
          I. Pessa
          \inst{1}
          .
          }

   \institute{Leibniz-Institut f\"ur Astrophysik Potsdam (AIP),
        An der Sternwarte 16, 14482 Potsdam, Germany
        \and
        Canada-France-Hawaii Telescope, 65-1238 Mamalahoa Highway, Kamuela, HI 96743, USA
        \and
        Aix Marseille Univ, CNRS, CNES, LAM, Marseille, France
        \and
        Institut de Recherche en Astrophysique et Planétologie (IRAP), Université de Toulouse, CNRS, UPS, CNES, Toulouse, France
        \and
        Centre de Recherche Astrophysique de Lyon UMR5574, Univ Lyon1, ENS de Lyon, CNRS, F-69230 Saint-Genis-Laval, France 
        \and
        Max Planck Institute for Astronomy, K\"{o}nigstuhl 17, 69117 Heidelberg, Germany
        }

   \date{Received 27 February 2024 / accepted 11 April 2024}
 
  \abstract
   {We quantify the stellar rotation of galaxies by computing the $\lambda_{R}$ parameter, a proxy for the stellar angular momentum in a sample of 106 intermediate-redshift galaxies (0.1 $<$ z $<$ 0.8). The sample is located in the CANDELS/GOODS-S and CANDELS/COSMOS regions, and it was observed by various MUSE surveys. We created spatially resolved stellar velocity and velocity dispersion maps using a full-spectrum fitting technique, covering spatially $\sim$2$R_{e}$ for the galaxies. The sample spans stellar masses from $\sim$10$^{7.5}$ M$_{\odot}$ to 10$^{11.8} $ M$_{\odot}$ with star formation rates (SFRs) from log$_{10}$(SFR) $\approx$-3 M$_{\odot}$yr$^{-1}$ to $\approx$1.7 M$_{\odot}$yr$^{-1}$ over a range of 6 Gyr in cosmic time. We studied how the atmospheric seeing, introduced by the instrumental point spread function (PSF), affects the measured spin parameter, and we applied corrections when pertinent. Through the analysis of the $\lambda_{R}-\epsilon$ diagram, we note that the fraction of round and massive galaxies increases with redshift. We did not measure any galaxy with $\lambda_{R}$ < 0.1 in the sample, and we found only one potential (but uncertain) low-mass slow rotator at z $\sim0.3$, more similar to the z=0 low-mass slow rotators characterized by counter-rotation than to massive ellipticals. Moreover, we do not see an evident evolution or trend in the stellar angular momentum with redshift. We characterized the galaxy environment using two different indicators: a local estimator based on the Voronoi tesselation method, and a global estimator derived by the use of the friends-of-friends (FoF) algorithm. We find no correlation between the environment and $\lambda_{R}$ given that we are not probing dense regions or massive galaxy structures. We also analysed the kinematic maps of the sample finding that about 40\% of the galaxies are consistent with being regular rotators (RRs), having rotating stellar discs with flat velocity dispersion maps, while $\sim20\%$ have complex velocity maps and can be identified as non-regular rotators in spite of their $\lambda_{R}$ values. For the remaining galaxies the classification is uncertain. As we lack galaxies with $\lambda_{R}$ < 0.1 in the sample, we are not able to identify when galaxies lose their angular momentum and become slow rotators within the surveyed environments, area, and redshift range.}

   \authorrunning{Constanza Mu\~{n}oz L\'{o}pez et al.}
   \maketitle
%
%-------------------------------------------------------------------
\section{Introduction}

   Studying the channels of galaxy evolution and how they are related to the observed stellar angular momentum requires taking into account several galaxy properties, such as  morphology, metal content, stellar age, the history of mass assembly, and their dependence on the environment. In this work we investigate the rotational support in galaxies at different cosmic times and environments with the purpose of determining how galaxy stellar angular momentum evolves.
    
   The orbital design of a galaxy is echoed by its morphology. Low angular momentum galaxies exhibit stellar systems with high random motions and tend to present spheroidal or circular shapes \citep{2016ARA&A..54..597C, 2020MNRAS.495.1958W}. On the other hand, high angular momentum galaxies are dominated by stars with an ordered rotation located in discs. It is not clear, however, when or how galaxies lose their angular momentum as we lack observations of slow rotators (SRs) at z > 0. At low redshifts (z $\sim$ 0) \citep{2009ApJ...699..105C, 2019ApJ...874...67B} and at higher redshifts (z $>$ 0.1)  \citep{2011ApJ...742...96W, 2017ApJ...838...19W} there is evidence that the structure (e.g. light distribution, S\'{e}rsic index) and the star-forming state of galaxies are linked, suggesting that quenching mechanisms and morphology are closely connected. Using integral field spectroscopic (IFS) surveys, the kinematics of high-$z$ ($z =$ 1-3) galaxies have been analysed mainly via their ionized gas content (KMOS$^{3D}$, \citealp{2015ApJ...799..209W}; KROSS, \citealp{2016MNRAS.457.1888S}). At these redshifts, it is expected that most galaxies are gas-rich and have high star formation rates (SFRs) as redshift coincides with the peak of SFR density at around $z \approx$ 1.9 or $\sim$3.5 Gyr after the Big Bang \citep{2014ARA&A..52..415M}.

   Focusing on the stellar kinematics, numerical simulations predict that the slow-rotating galaxy population starts to build up around z$\sim$2 and become more prominent towards z$\sim$0  \citep{2011MNRAS.417..845K, 2013MNRAS.429.1258D}. \cite{2018MNRAS.480.4636S} used the Cosmological hydrodynamical Magneticum Pathfinder simulation and found that slow-rotating galaxies become statistically significant below $z =$ 1 with a gradual increase. Moreover, \citep{2018MNRAS.476.4327L} using EAGLE simulations, showed that galaxies experiencing successive minor mergers are more prone to lose their spin and become SRs. This hierarchical assembly through gas-poor minor merging plays an important role in building the population of nearby elliptical galaxies \citep{2009ApJ...699L.178N, 2015MNRAS.449..361W}. The expectation is that as early-type galaxies (ETGs, i.e. ellipticals and S0s) continue to  grow in mass and size, their kinematics evolve as well, and they  transition from rotation-supported to pressure-supported systems \citep{2011MNRAS.416.1680C, 2013ApJ...771...85V}. 

   Early-type galaxies in the nearby Universe can be classified, using IFS kinematics, as slow or fast rotators \citep{2007MNRAS.379..401E, 2011MNRAS.414..888E} based respectively on whether they present or not regular rotation \citep{2011MNRAS.414.2923K}. Slow-rotating galaxies are thought to be the final products of galaxy evolution representing around one-fifth of the stellar mass in the current Universe \citep{2011MNRAS.417..845K, 2013MNRAS.429.1258D}. This type of galaxy is characterized by showing little to no rotation, being typically massive (M$_{\star}\sim$10$^{11}$ M$_{\odot}$), presenting little evidence for recent or ongoing star formation; they are frequently found in dense environments, and lack stellar-ordered motion. They are also galaxies with complex orbital structures, different families of orbits, including galaxies that have zero mean angular momentum. On the other hand, fast-rotators present regular stellar velocity fields, discy isophotes, consistent with disc-like rotation or structures \citep{2011MNRAS.414.2923K}, and are galaxies with important dispersion support. In fast-rotators (spirals and S0 galaxies), the regular stellar rotation happens in discs. However, some ellipticals are flattened by rotation, where the ordered motion dominates over the random motion, and while they do not have thin discs they have simple orbital structures \citep{2013MNRAS.432.1768K}. Various studies \citep{2011ApJ...739...45F, 2013ApJ...768...74T, 2014ApJ...792L...6V, 2017ApJ...834...18B} have shown that at z$\sim$2 massive galaxies, including  quiescent galaxies, exhibit significant angular momentum, which is in contrast with galaxies of similar mass in the local Universe. 

   This scenario suggests an evolution in the angular momentum from high to low redshifts. The mechanisms behind the galaxy's stellar angular momentum evolution through cosmic time and the SR formation scenario are still open questions. One possibility is to consider the hierarchical growth of galaxies in clusters, using galaxy merger trees \citep{2012MNRAS.423.1277D}, where the progenitor cores of SRs are formed at high redshift at the centre of dark matter overdensities. Within the hierarchical buildup of galaxies and cluster  hypothesis, SRs sink towards the centre of the resulting group or cluster of galaxies, where they merge and form a more massive SR. {\citealp{2022MNRAS.509.4372L}, using EAGLE simulations, observed that the   SRs that experienced minor or major mergers are more prone to be triaxial systems. On the other hand, SRs that do not experience mergers tend to be more compact and are quenched later compared to the other SRs}. Another formation scenario is that quenching processes destroy the organized rotation of galaxies. As the galaxy quenches, the angular momentum diminishes, or as an alternative the decreasing organized rotation itself is responsible for quenching the galaxy \citep{2009MNRAS.398..898H}. Moreover, morphological quenching can be effective, where the growth of the bulge is linked to the decrease in the angular momentum, and to the quenching of the star formation \citep{2009ApJ...707..250M}. Thus, characterizing the kinematics of galaxies is crucial to understanding their formation and evolution.
   
   Local surveys (e.g. ATLAS$^{3D}$, \citealp{2011MNRAS.413..813C}; CALIFA, \citealp{2012A&A...538A...8S}; SAMI, \citealp{2012MNRAS.421..872C}; MaNGA, \citealp{2015ApJ...798....7B}) have studied the stellar angular momentum in galaxies and how this relates to other galaxy properties. Using the ATLAS$^{3D}$ survey, \cite{2011MNRAS.414..888E} found that the fraction of SRs increases strongly with stellar mass for a sample of $\sim$260 nearby ETGs, a finding confirmed by all subsequent studies. Likewise, \cite{2011MNRAS.416.1680C}, found that the slow rotator population increases in denser environments. On the other hand, \cite{2017ApJ...844...59B}, using the stellar mass-selected SAMI survey, which included both early- and late-type galaxies in various environments, found little or no relationship between the spin parameter and the environmental local overdensity once the stellar mass is accounted for. In galaxy groups, \cite{2017ApJ...851L..33G} used MaNGA data and found that the slow rotator fraction is only due to the stellar mass, not the environment. A similar result was found in \citet{2017ApJ...851L..33G}. Studying $\sim$1800 early- and late-type galaxies, \citet{2021MNRAS.508.2307V} found that the stellar mass (> 10$^{9.5}$M$_{\odot}$) is the primary driver of angular momentum loss, but the environment has a significant role \citep[see also][]{2020MNRAS.495.1958W}: galaxies in high-density environments have lower angular momentum compared with low-density environments. {In a study with SAMI galaxies, \citet{2024arXiv240206877C} built a regression model to determine whether a galaxy is a slow or fast rotator from a variety of galaxy properties. They found that the galaxy's environment at fixed stellar mass, star formation rate, size, and ellipticity is not a useful indicator, in their model to discern the kinematic morphology of galaxies. This results from the increase in the fraction of slow rotators in denser environments, which is a consequence of quenched massive elliptical galaxies being mostly found in dense environments. Furthermore, \citet{2024MNRAS.528.5852V}, using SAMI galaxy survey, studied the drivers of the spin parameter ($\lambda_{R}$) and found that for low- and intermediate-mass galaxies, the light-weighted stellar age or specific star formation rate is the primary parameter affecting $\lambda_{R}$, while the environment affects it indirectly.}
%--------------------------------------------------------------------
    \begin{figure*}
    \centering
       \includegraphics[width=17cm]{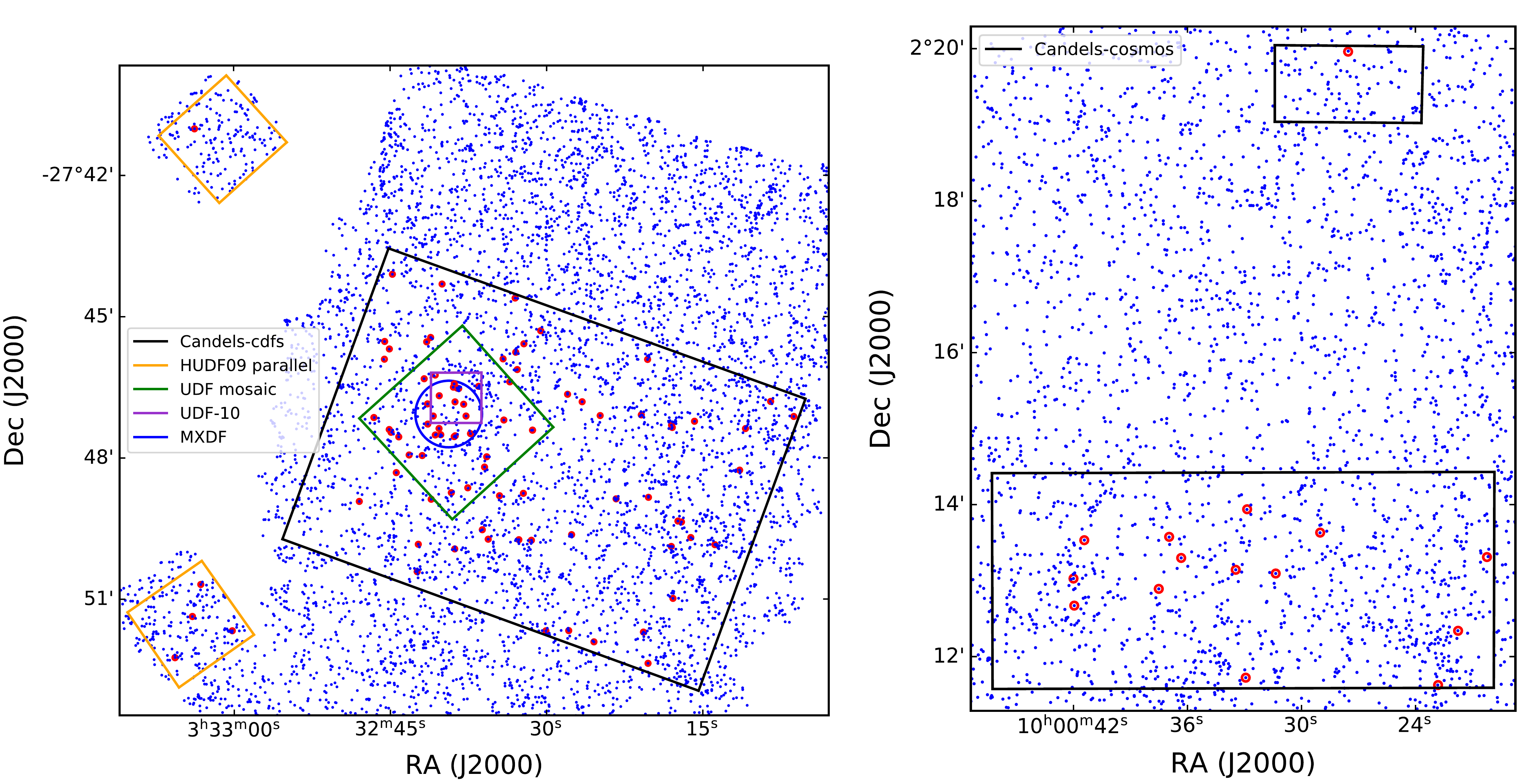}
         \caption{Layout of the MUSE-Wide regions with the galaxy sample represented by red circles. $\textit{Left:}$ blue dots correspond to HST/WFC3 objects from the 3D-HST survey in the GOODS-S region brighter than 24 mag in the F160W band. The black, orange, green, purple, and blue contours  approximately delimit the candels-cdfs, HUDF09 parallel, UDF mosaic, UDF-10, and MXDF regions, respectively. $\textit{Right:}$ Sample galaxies (red symbols) in the candels-cosmos fields. The black line encloses the area covered by the MUSE-Wide survey. The red and blue symbols are the same as in the \textit{left} panel. For more details about the fields, see Section \ref{sec:data}.}
         \label{fig:musewide_fields}
    \end{figure*}

%--------------------------------------------------------------------    
   Understanding the evolution of the stellar angular momentum is not a trivial task. We would like to know the epoch and the driving mechanisms (mass, environment, star formation quenching) of slow-rotator formation. So far, we do not have a clear picture of the behaviour of the stellar rotation of galaxies from observations at higher and broader redshift ranges. The stellar kinematics of intermediate-redshift galaxies have been analysed over the past in several studies. \cite{2007ApJ...668..738V} analysed the velocity and velocity dispersion profiles of 30 cluster galaxies at z$\sim$0.5 through the inspection of Keck/LRIS absorption lines, and also the use of dynamical models. They found that the brightest galaxies in the sample presented overly low rotation to account for their flattening, possibly due to the high fraction of misclassified S0 galaxies. \cite{2008ApJ...684..260V} used 25 field early-type galaxies with 0.6 < $z$ < 1.2, and found no evolution in the fraction of early-type fast rotators compared to local galaxy samples. Moreover, \cite{2018ApJ...858...60B}, utilizing long-slit spectroscopic data from the LEGA-C survey, studied the stellar rotation curves and velocity dispersion profiles of $\sim$100 massive quiescent galaxies at $z =$ 0.6-1, and found that the rotational support decreases with increasing mass, a similar trend to early-type galaxies in the local Universe. 
    
   The first work that used the advantage of integral field units (IFUs) to analyse the stellar kinematics of intermediate-redshift galaxies was \cite{2017A&A...608A...5G}. They used 17 galaxies observed with the Multi Unit Spectroscopic Explorer \citep[MUSE, ][]{2010SPIE.7735E..08B} located at the Very Large Telescope (VLT), with 0.2 < $z$ < 0.8 to study the gas and stars. They found that the kinematics of local disc galaxies were already set 4-7 Gyr ago. Following this route, our aim is to use the mean velocity and the velocity dispersion maps, also based on MUSE observations, to measure the stellar angular momentum of intermediate-redshift galaxies in a larger sample. 
   We aim to study a specific range in the redshift space ($z < $ 1) making use of the long spectral coverage of MUSE, as Ca II H and K lines are redshifted to 7936 and 7868 $\AA$ at z=1, respectively. In particular, we present the kinematic analysis of a sample composed of 106 galaxies with spectroscopic redshift ranging from 0.1 to $\sim$0.8 and stellar masses in the range of 10$^{7.5}$--10$^{11.8}$ M$_{\odot}$.

   The paper is structured as follows. In Section \ref{sec:data} we describe the characteristics of the data that were used in our work. In Section \ref{sec:sample} we provide details about how the galaxy sample was constructed. In Section \ref{sec:kinematics} we explain the method used to measure the stellar angular momentum in the sample, and how this is affected by the atmospheric seeing. In Section \ref{sec:results} we discuss the main results of our work, which includes the analysis of kinematic maps, morphology, stellar parameters of the galaxies, and some environmental indicators retrieved to characterize the environment in the sample. Throughout the paper we adopt a flat Universe, H$_{0}$ = 70 km s$^{-1}$ Mpc$^{-1}$, $\Omega_{\Lambda}$ = 0.7 cosmology.
   
\section{Data sets}\label{sec:data}

\subsection{MUSE observations}\label{sec:MUSE}

    We used data sets from the blind 3D spectroscopic survey MUSE-Wide \citep{2019A&A...624A.141U}, which targets 100 fields and/or pointings in the CANDELS/GOODS-S and CANDELS/COSMOS regions. The survey is the wide and shallow counterpart to the blind MUSE-Deep surveys in the Hubble Ultra Deep Field (HUDF, \citealp{2017A&A...608A...1B, 2023A&A...670A...4B}) following a `wedding-cake' observing depth approach, in which MUSE-Wide covers in total $100 \times 1$ arcmin$^{2}$ fields for one hour, while MUSE-Deep covers 9 x 1 arcmin$^{2}$ fields for 10 hours, $1 \times 1$ arcmin$^{2}$ field to 31 hours, and 141 hours on a circular field with $1^{\prime}$ diameter. All MUSE observations were taken in the wide-field nominal wavelength (480--930 nm) mode and not using adaptive optics (AO), except in one case, as noted below. 

        The details about the observing strategy and the data reduction process can be found in \cite{2017A&A...608A...1B} and \cite{2019A&A...624A.141U}, with the exception of some modifications performed afterwards to obtain a better version of the data cubes: (i) The data was reduced with  version 2.0 of the MUSE Data Reduction Pipeline \citep{2020A&A...641A..28W}, which implements a much improved deep-field autocalibration; (ii) sky-subtraction was improved using a new version of the Zurich Atmospheric Purge (ZAP; \citep{2016MNRAS.458.3210S}) algorithm; (iii) a superflat was created for each exposure to account for the defects and `dark spots' in the slicer stack transitions; (iv) some small changes to the final cube combination regarding edge masking and sigma clipping were implemented; and  (v) six fields were re-observed in 2017 in order to  replace the old bad-quality fields. A new data release of these v2.0 MUSE-Wide data is expected in late 2024.

    Figure \ref{fig:musewide_fields} shows the layout of the MUSE regions in the GOODS-S and COSMOS fields. The characteristic of the regions are described in the following paragraphs. 
    %%%%%%%%%%%%%%%%%%%%%%%%%%%%%%%%%%%%%%%%%%%%%%%%%%%%%%%
    \begin{figure*}
    \centering
       \includegraphics[width=18cm]{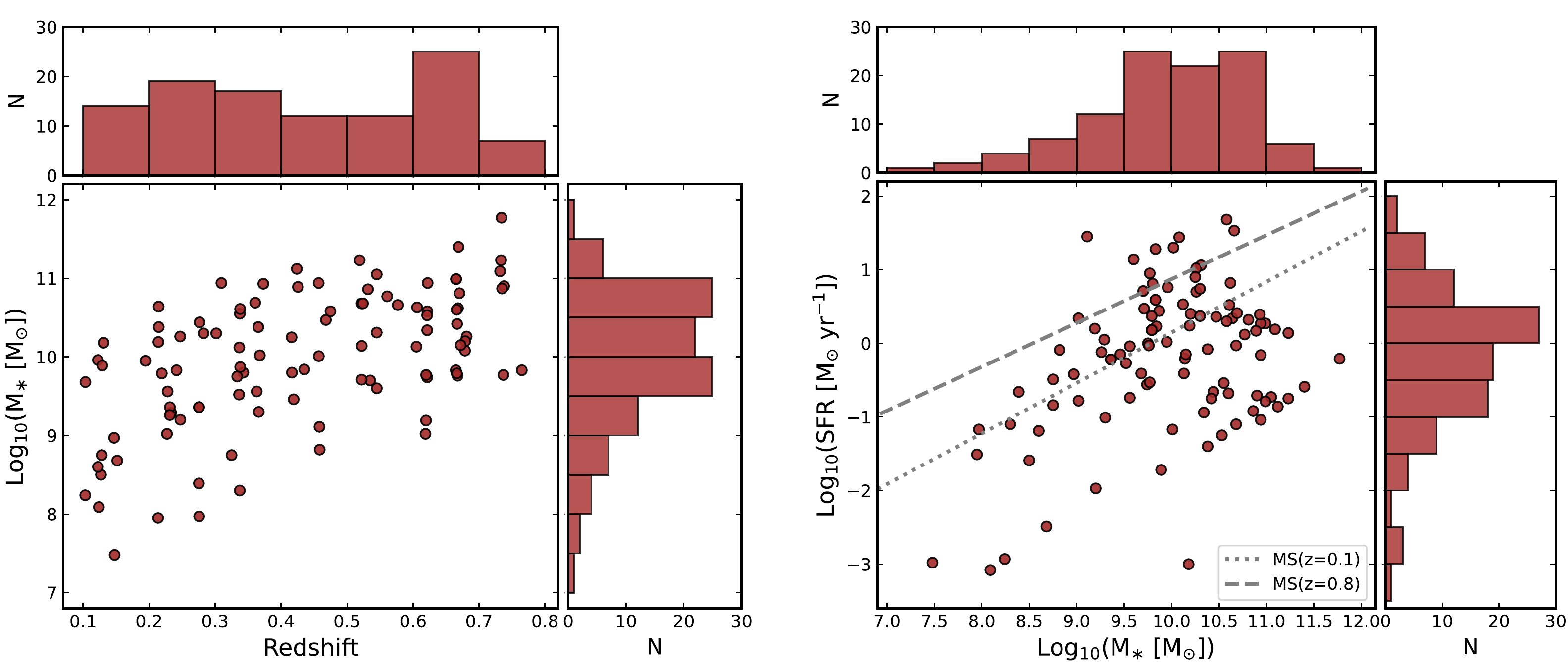}
         \caption{General properties of the galaxy sample. $\textit{Left:}$ Stellar mass distribution as a function of the galaxy spectroscopic redshift. $\textit{Right:}$ Star formation rate as a function of the stellar mass for the sample galaxies. The dotted and dashed grey lines indicate the main sequence (MS) for an age of the Universe respectively of 12.4 Gyr (z=0.1) and 6.8 Gyr (z=0.8), given by Eq. 1 of \cite{2012ApJ...754L..29W}.}
         \label{fig:red_mass_sfr}
    \end{figure*}
%%%%%%%%%%%%%%%%%%%%%%%%%%%%%%%%%%%%%%%%%%%%%%%%%%%%%%% 
\subsubsection{CANDELS-CDFS region}\label{sec:candels}

    Of the 100 MUSE-Wide pointings, 60 of them are named `candels-cdfs' and are overlaying the CANDELS/GOODS-S region. Each field covers an area of $1' \times 1'$ with a position angle PA of 340$^{\circ}$, achieving an exposure time of 1 h. Candels-cdfs fields are indexed from 1 to 62 according to the observing queue.\footnote{ESO programme identification: 099.A-0060(A)} The MUSE observations were carried out between October 2014 and March 2016.

\subsubsection{HUDF09 parallels region}\label{sec:hudf}

    The MUSE-Wide survey also comprises eight MUSE fields in two different mosaics of $4' \times 4'$, designated as HUDF09-1 and HUDF09-2. The pointings are located in the Hubble Ultra Deep Field (HDUF) parallel fields \citep{2011ApJ...737...90B} reaching a depth of 1 h.\footnote{ESO programme identification: 094.A-0205(B)} To match the CANDELS-Deep field layout, the pointings were observed with a PA of 42$^{\circ}$ for HUDF09-1, and 35$^{\circ}$ for HUDF09-2. The data were taken in September 2014.

\subsubsection{UDF deep mosaic region}\label{sec:cudf-mosaic}

    The UDF deep mosaic \citep{2017A&A...608A...1B} covers around 90$\%$ of the HDUF region with nine 1' x 1' MUSE pointings, denoted from UDF-01 to UDF-09. The mosaic is orientated with a PA of -42$^{\circ}$. Some of the fields overlap with the candels-cdfs regions (Section \ref{sec:candels}), but the integration time is significantly greater, with an exposure time of 10 h. The observations were performed in September 2014.\footnote{ESO programme identification: 094.A-0289(B)} 

\subsubsection{UDF-10 region}\label{sec:udf-10}

    The Ultra Deep Field-10 (UDF-10) located in the Hubble Ultra Deep Field (HUDF) \citep{2006AJ....132.1729B} is composed of a unique pointing covering an area of 1 x 1 arcmin$^{2}$ with a PA of 0$^{\circ}$. The field reaches a depth of $\approx$31 h of integration time, and it was observed between September 2014 and December 2015. A similar analysis to that presented in this paper was performed previously by \cite{2017A&A...608A...5G} with galaxies belonging to this field, which are re-used in this work. 
 
\subsubsection{MXDF region}\label{sec:mxdf}

    The MUSE eXtremely Deep Field \citep[MXDF;][]{2023A&A...670A...4B} is a  circular field of 1 arcmin in diameter located in the HUDF extremely deep field region. The observations were carried out using AO, reaching 141 h of integration time.
    %%%%%%%%%%%%%%%%%%%%%%%%%%%%%%%%%%%%%%%%%%%%%%%%%%%%%%%
    \begin{figure*}
    \centering
       \includegraphics[width=18cm, height=10.5cm]{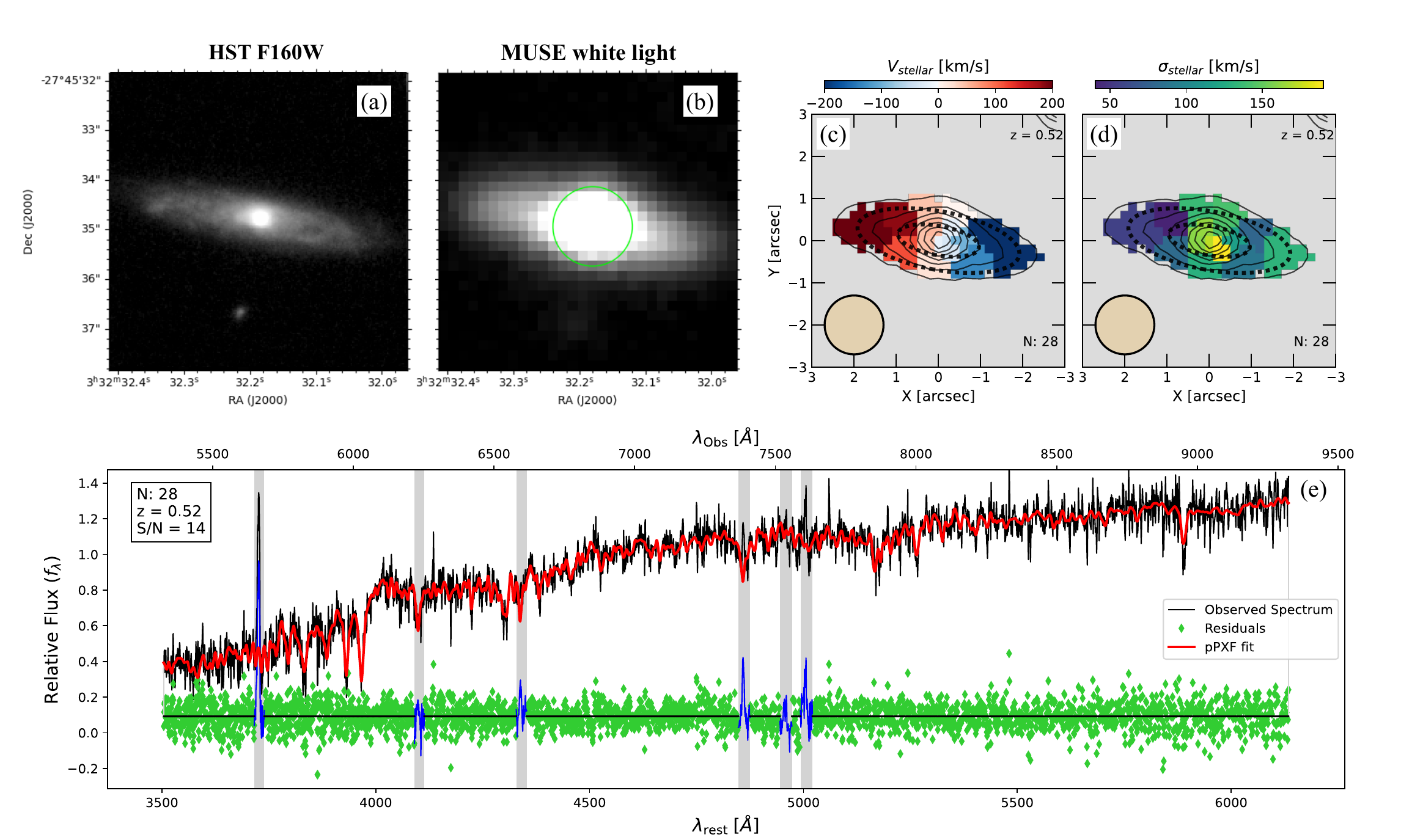}
         \caption{Example of the kinematics analysis for the galaxy with galaxy 28 in the sample. \textit{Panel (a):} HST image of the galaxy in the F160W band. \textit{Panel (b):} MUSE white light image of the galaxy. The green circle corresponds to the effective radius of the galaxy. \textit{Panels (c) and (d):} Resolved stellar velocity and velocity dispersion maps of the galaxy. The maps are Voronoi binned. The contours are isophotes of the surface brightness plotted with black continuous lines. The number $N$ in the lower right part of each panel denotes the galaxy index according to Table \ref{tab:sample_table}. The velocity and velocity dispersion ranges are shown in different colours (see colour bar at the top of each  panel). The inner dashed black line correspond to the ellipse computed using 1$R_{e}$ (half-light ellipse) and the outer ellipse by taking 2$R_{e}$. \textit{Panel (e):} pPXF fit of the spectrum within the effective radius of the galaxy. The black line indicates the observed spectrum, the red line corresponds to the best fit, the green diamonds are the residuals of the fit, and the shaded grey regions correspond to the masked emission lines. The galaxy $N$ number, the redshift, and the signal-to-noise ratio are indicated in the upper left corner of the panel.}
         \label{fig:example_gal}
    \end{figure*}
%%%%%%%%%%%%%%%%%%%%%%%%%%%%%%%%%%%%%%%%%%%%%%%%%%%%%%%
\subsubsection{CANDELS-COSMOS region}\label{sec:cosmos}

    The CANDELS-COSMOS fields from the MUSE-Wide survey consist of 22 pointings covering an area of $\approx$22 arcmin$^{2}$ located in the COSMOS region \citep{{2007ApJS..172....1S}}. The region reaches a depth of 1 h of integration time.\footnote{ESO programme identification: 097.A-0160(A)} The pointings numbered from 1 to 21 are displayed in a rectangular mosaic with a PA of 0$^{\circ}$, plus two other fields, 58 and 59, which are located further to the north. These regions are named `candels-cosmos' with the respective field number. 

\subsection{HST observations}\label{sec:hst}

    We used photometric catalogues from the 3D-HST (\citealp{2011ApJ...743L..15V, 2012ApJS..200...13B, 2014ApJS..214...24S}) near-infrared survey which cover $\approx$900 arcmin$^{2}$ in the AEGIS, COSMOS, GOODS-South, and UDS fields, therefore covering the full MUSE-Wide and Muse-Deep area. The data contain WFC3 image mosaics of all five CANDELS/3D-HST \citep{2011ApJS..197...35G} fields in various photometric filters. The observations and analysis are described exhaustively in \cite{2014ApJS..214...24S}. Some parameters from the catalogues, such as magnitudes and aperture size, were utilized to create  our galaxy sample and, later, to characterize it.

    The photometric catalogues were produced using point spread function (PSF)-matched aperture photometry. For all the fields the detection image is a noise-equalized mix of F125W, F140W, and F160W filters, with a final pixel scale of 0''.06 pixel$^{-1}$. The 3D-HST stellar population parameters, such as stellar masses, star formation rates, ages, and dust extinctions, were estimated using version 0.8d of the code FAST \citep{2009ApJ...700..221K}.

\section{Galaxy sample}\label{sec:sample}

    The sample members were selected based on photometric and spectroscopic parameters. The approach adopted in the sample creation was focused on maximizing the number of galaxies that can be spatially resolved and is based on the recipe used by \cite{2017A&A...608A...5G}.  

    Despite the high number of detected galaxies in MUSE-Wide fields \citep{2019A&A...624A.141U}, most of them are   outside of the target redshift range, lack sufficient S/N, or are not spatially large enough to be included in our sample. These aspects are the main constraints in shaping the final sample, and they will be addressed in this section.
    
\subsection{Selection criteria}\label{sec:selection} 

    Using the MUSE catalogues, we first selected galaxies with I$_{F160W} \leq$ 24 mag and with spectroscopic redshift z $\leq$ 1.2. For z $\geq$ 1.2, all absorption lines useful for measuring stellar kinematics leave the MUSE wavelength range. Then we extracted a global galaxy spectrum containing half of the galaxy's light. The effective radius measurements were taken from \cite{2014ApJS..214...24S}. The global spectrum corresponds to the average spectrum from all the spaxels enclosed by the effective radius. A primary Penalized PiXel-Fitting method \cite[pPXF,][]{2004PASP..116..138C,2017MNRAS.466..798C} was used to get a global estimator of the galaxy S$\slash$N (using the full wavelength range) for $\sim$1500 objects. The fitting method used is described in Section \ref{sssec:method}. The galaxy sample was formed of galaxies with a global S$\slash$N $\gtrsim$ 10. This S$\slash$N cut was used based on simulations performed by \cite{2017A&A...608A...5G}, which show that for S$\slash$N$\sim$8 pixel$^{-1}$ it is possible to recover velocity dispersion down to 40 km$\slash$s with an acceptable error of 10 km$\slash$s. 
    
    To increase and homogenize the spectral S$\slash$N of the sample galaxies, we spatially binned them using the code based on Voronoi tessellation developed by \cite{2003MNRAS.342..345C}. The noise input comes from the MUSE variance data cubes, products of the MUSE pipeline. We explored how estimating the S$\slash$N from different wavelength ranges affected the Voronoi binning, and therefore the number of bins that we can achieve per galaxy. The rest-frame wavelength range selected is either 4150-4350 or 4500-4700 $\AA$, and only pixels with a S$\slash$N $\geq$ 1 are considered for the binning process. The selected intervals are free of emission lines with only weak absorption lines, and thus representative of the continuum level; they are also the ranges that maximize the number of galaxies that can be resolved in multiple spatial bins. We slightly modified the \cite{2017A&A...608A...5G} restriction, and required, for each galaxy, that there are at least five Voronoi bins with S$\slash$N $\geq$ 8. To ensure that we did not neglect any galaxy, a visual inspection of all galaxies within a MUSE field was performed using HST images in different bands (F814W, F850LP, F160W) depending on the availability. 

    The final sample is composed of 106 galaxies: 53 in the candels-cdfs fields, 15 in the candels-cosmos fields, 16 in the udf-mosaic fields, 13 in the MXDF region, 5 in the HUDF09 parallels, and 4 in the UDF-10 fields. In Appendix \ref{sec:table_maps} and Table~\ref{tab:sample_table} we list the properties of each galaxy in the sample.

\subsection{Global properties}\label{sec:global-properties}

    As outlined above, the sample is composed of 106 galaxies distributed in the MUSE fields. Figure \ref{fig:musewide_fields} shows the positions of the galaxies over the GOODS-S and the COSMOS regions. The sample galaxies are more concentrated towards fields where the integration time is higher (i.e. the MXDF field). The number of galaxies in the sample is affected mainly by the degradation in the apparent magnitude of galaxies with the increase in redshift, and the galaxy's angular size which limits the number of bins achieved with a desired S/N, and as a consequence regulates the number of galaxies in the sample.

    Of the 106 galaxies, the faintest one reaches an apparent magnitude of $\sim$23.6 mag, while the brightest galaxy has a magnitude $\sim$16.6 mag in the HST F160W band. Figure \ref{fig:red_mass_sfr} shows the spectroscopic redshift range of the sample, which spans from $\sim$0.1 to $\sim$0.76. The sample has a similar number of galaxies between z$\sim$0.1 and z$\sim$0.6. The redshift distribution presents a clear peak at 0.6 $\leq$ z $\leq$ 0.7, where galaxies have masses $\geq$10$^{9.5}$ M$_{\odot}$. The derived stellar masses through SED fitting of the sample ranges from $\sim$ 10$^{7.5}$ M$_{\odot}$ to $10^{11.8} $ M$_{\odot}$ with SFRs from log$_{10}$(SFR) $\approx$ -3 M$_{\odot}$yr$^{-1}$ to $\approx$ 1.7 M$_{\odot}$yr$^{-1}$. The mass distribution presents a broad peak, with a similar number of galaxies between 10$^{9.5}$ M$_{\odot}$ and 10$^{11}$ M$_{\odot}$. These population parameters (SFR and stellar masses) were determined with the FAST code as part of the 3D-HST programme \citep{2014ApJS..214...24S}, and are listed in Table \ref{tab:sample_table}.
    MXDF galaxies represent $\sim$12$\%$ of the sample in number while only covering $\sim$1$\%$ of spatial area. The MXDF data are much deeper than the other fields, not only allowing more galaxies to be caught at higher redshifts, but also less massive galaxies at low redshifts (z $\sim$ 0.1-0.3).
    
        The right panel of Fig. \ref{fig:red_mass_sfr} shows that the galaxy sample covers the star formation main sequence in the redshift range z $\sim$ 0.1 - 0.8. While the majority of the galaxies are star-forming, some galaxies are reaching the passive or quenched region, as seen in the lower part of the diagram.

\section{Stellar kinematics}\label{sec:kinematics}

\subsection{Method}\label{sssec:method}

    As mentioned earlier, to extract the resolved stellar kinematics of the galaxy sample, we used the pPPX algorithm, a method for extracting the stellar and gas kinematics, and the stellar populations of galaxies by fitting a combination of templates from a stellar library to an observed spectrum. We focused on the recovery of the two first-order moments of the line-of-sight velocity distribution (LOSVD), the radial mean velocity ($V$), and the velocity dispersion ($\sigma$). As we are limited by relatively low S/Ns and spatial resolution, we did not recover  higher-order Gauss-Hermite moments \citep{1993ApJ...407..525V, 1993MNRAS.265..213G}. 
    
    For the kinematics extraction, we used a gap-free subset of 195 templates from the Indo-US stellar library \citep{2004ApJS..152..251V}. We  selected  this library because of  its constant spectral resolution of 1.35 $\AA$ full width at half maximum (FWHM) \citep{2011A&A...531A.109B} over its entire wavelength range (i.e. 3460 - 9464 $\AA$), which is finer than the MUSE line-spread function (LSF) FWHM  even for the galaxies with the highest redshifts in our sample. 
    
    Before the fitting started, the stellar templates were convolved to a wavelength dependent MUSE LSF resolution as derived in \cite{2017A&A...608A...5G}. We masked several gas emission lines: [OII] $\lambda \lambda$3726,3728, H$\delta$ $\lambda$4101, H$\gamma$ $\lambda$4340, H$\beta$ $\lambda$4861,  [OIII] $\lambda \lambda$4958,5006,  [OI] $\lambda$6300,  [NII] $\lambda$6548, 6583, H$\alpha$ $\lambda$6563, and [SII] $\lambda \lambda$6716, 6730. We spectrally re-binned both the stellar templates and the MUSE spectra. The template spectra were logarithmically rebined to a velocity scale two times smaller ($\approx$ 27 km s$^{-1}$ pixel$^{-1}$) than the MUSE galaxy spectra ($\approx$ 55 km s$^{-1}$ pixel$^{-1}$), as prescribed in \cite{2017MNRAS.466..798C}. To fit the velocity and velocity dispersion of the galaxies we used additive polynomials of 12th order and a zero degree multiplicative polynomial.
    %We assume a constant noise per pixel of 0.0047 to describe the uncertainties in the galaxy spectrum.
    
    A fitting example can be seen in the bottom panel of Fig. \ref{fig:example_gal}. We present the pPXF fit and the residuals for the integrated spectrum of one sample galaxy within its effective radius, which is shown as a green circle in the upper left (b) panel, as well as the resulting mean velocity and velocity dispersion maps.
    
    We also computed the errors in the extracted velocity and velocity dispersion per galaxy bin. Firstly, we performed an initial pPXF fit per Voronoi spatial bin obtaining the fitted spectrum, and the difference between the observed and modelled bin spectrum (i.e. the residuals of the fit). Then for each Voronoi bin, we calculated the mean and standard deviation of the residuals. Assuming that these follow a normal distribution, we added random Gaussian noise to the modelled spectrum of the bin and produced 100 new spectra per bin. After fitting these new spectra, we obtained a distribution of velocities and velocity dispersions per bin, and by taking the standard deviation from both distributions, we obtained the uncertainties for the extracted velocities and velocity dispersions. 
 
\subsection{The spin parameter $\lambda_{R}$}\label{sec:lam}

    We derived the spin parameter $\lambda_{R}$, a proxy for the luminosity-weighted stellar angular momentum, to quantify the global rotational velocity structure of galaxies using the 2D spatial data. As we produced stellar kinematic maps for each galaxy in the sample, $\lambda_{R}$ is calculated using equations (5) and (6) of \cite{2007MNRAS.379..401E}:

    \begin{equation}
    \lambda_{R} \equiv \frac{\langle R \mid V \mid \rangle}{\langle R \sqrt{V^{2} +\sigma^{2}}\rangle} = \frac{\sum_{i=1}^{N} F_{i} R_{i} \mid V_{i} \mid}{\sum_{i=1}^{N} F_{i} R_{i} \sqrt{V^{2}_{i} +\sigma^{2}_{i}}}
    \label{eq:lam}
    .\end{equation} 
Here $F_{i}$, $R_{i}$, $V_{i}$, and  $\sigma_{i}$ are the flux, the distance to the centre of the galaxy, the mean stellar velocity, and the velocity dispersion of the $i$-th pixel, respectively. The centre of the galaxy is determined as the pixel with the highest integrated flux. The values for $V$ and $\sigma$ for a bin are calculated with the flux contribution of each pixel inside it. Thus, the individual $V_{i}$ and $\sigma_{i}$ values for a particular pixel correspond to the bin velocity and velocity dispersion. Moving to the original pixel space, the value for $F_{i}$ corresponds to the pixel original flux ($F_{i}$). In this way, each pixel keeps its own flux, while its velocity and velocity dispersion are inherited from the bin  it belongs to. Then $\lambda_{R}$ is normalized by $\sqrt{V^{2}_{i} +\sigma^{2}_{i}}$, which is proportional to stellar mass and implies that the spin parameter goes to unity when the mean stellar velocity ($V$) dominates. The summation is performed over $N$ pixels within a fixed aperture (see Section \ref{sec:ellip_aper}).

%%%%%%%%%%%%%%%%%%%%%%%%%%%%%%%%%%%%%%%%%%%%%%%%%%%%%%%
    \begin{figure}
    \centering
       \includegraphics[width=9cm, height=6.5cm]{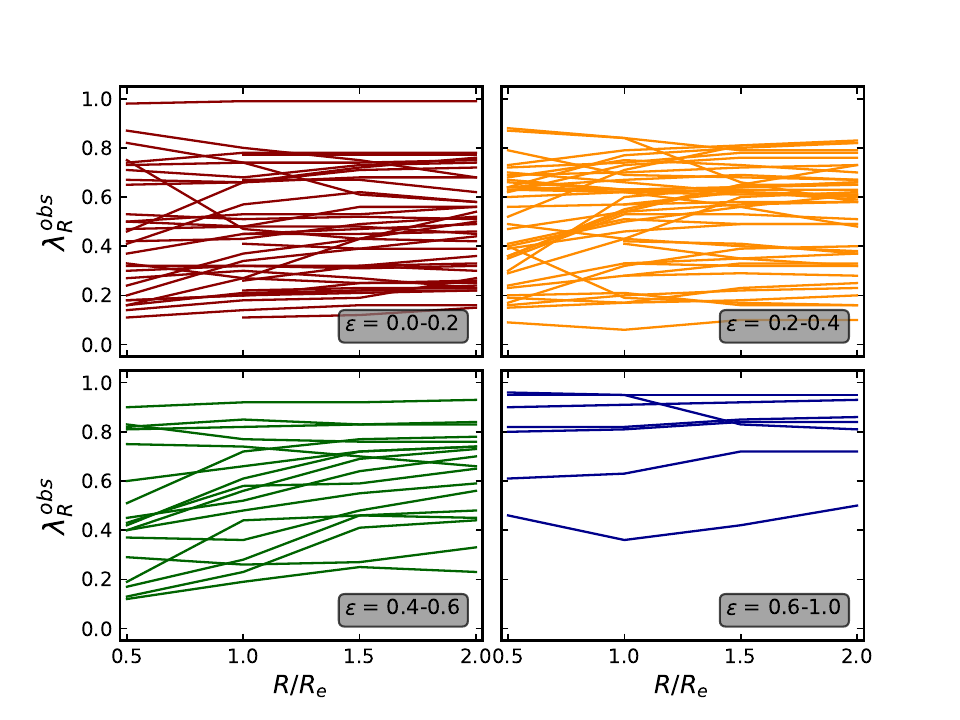}
         \caption{Observed stellar angular momentum profile for the sample galaxy for different aperture radii (0.5, 1.0, 1.5, and 2$R_{e}$) relative to the effective radius. The values are not seeing-corrected. The colours indicate the galaxy ellipticity $\epsilon$, which is also indicated in each panel.}
         \label{fig:lam_dif_aper}
    \end{figure}
%%%%%%%%%%%%%%%%%%%%%%%%%%%%%%%%%%%%%%%%%%%%%%%%%%%%%%%

    We applied a bootstrapping algorithm to the galaxy spectra to compute the errors in $\lambda_{R}$. As we wanted to estimate the errors in the angular momentum, we needed to obtain the errors in $V$ and $\sigma$ per bin. To obtain the error in the angular momentum, we created 1000 new velocity and velocity maps with random values between the originals ($V_{i}$, $\sigma_{i}$), and their respective bin errors. Finally, using Equation \ref{eq:lam} we calculated the angular momentum for all the new maps. Taking the standard deviation of the distribution of $\lambda_{R}$ values for the 1000 simulated realizations, the error in the observed $\lambda_{R}$ for each galaxy was obtained.

\subsection{Ellipticity and half-light ellipse} \label{sec:ellip_aper}

    In order to compute a measurement of the stellar angular momentum, which takes into consideration the galaxy's morphology, we needed to take into account its shape. Hence, we chose to calculate the spin parameter not only within the galaxy's half-light ellipse but also from 0.5 to 2 effective radii ($R_{e}$). 

        Therefore, to calculate the half-light ellipse, we used the galaxy circular aperture radius enclosing half the total flux ($R_{e}$), the position angle (PA), and the ellipticity ($\epsilon$), given by $\epsilon$ = $1 - \frac{b}{a}$, where $b$ and $a$ are the semi-minor and semi-major axes, respectively. All these parameters were obtained from the WFC3-selected catalogues of objects from the 3D-HST Survey \citep{2014ApJS..214...24S}. 
        Thus, the half-light ellipse is defined as an ellipse covering the same area ($A$ = $\pi R_{e}^{2}$) as a circle with radius $R_{e}$. The dashed black lines in panels (c) and (d) of Fig. \ref{fig:example_gal} show two regions: the half-light ellipse and an aperture computed taking $R$ = 2$R_{e}$.

        In Fig. \ref{fig:lam_dif_aper} we present the observed spin parameter computed for the galaxy sample at different apertures, from 0.5 to 2$R_{e}$ in four different panels according to the galaxy flattening. We see that the sample is dominated by round galaxies (low values of $\epsilon$). The radial $\lambda_{R}$ profile is flat, and therefore the observed kinematics do not change drastically between 0.5 and 2$R_{e}$ for most galaxies in the sample.
%%%%%%%%%%%%%%%%%%%%%%%%%%%%%%%%%%%%%%%%%%%%%%%%%%%%%%%
    \begin{figure*}
    \centering
       \includegraphics[width=17cm]{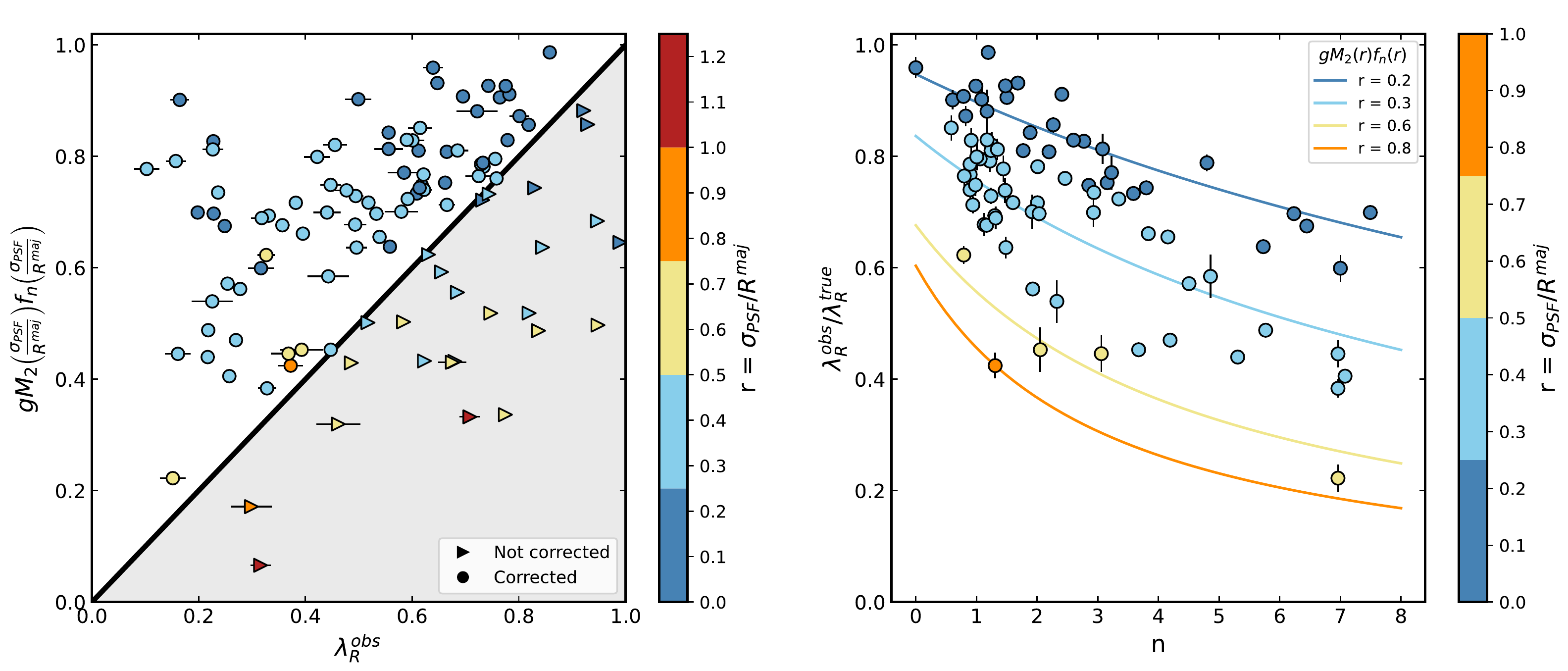}
         \caption{Corrections for the spin parameter in the galaxy sample. Both panels ($\textit{left}$ and $\textit{right}$) are coloured by the ratio ($r$) of the galaxy angular size to the PSF, given by the major-axis $R^{\rm maj}$ and $\sigma_{\rm{PSF}}$, respectively. The ratio $r$ is a parameter that quantifies how well-resolved a galaxy is. $\textit{Left:}$ Generalized Moffat and empirical $f_{n}$ functions as correction indicators from \cite{2018MNRAS.477.4711G} and their dependence on the observed (measured) stellar spin parameter, the galaxy angular size, and the width of the PSF. The shaded part of the plot indicates the region where the galaxy spin parameter correction is not physical, leading to  $\lambda_{R}^{\rm true} >1$. Thus, for galaxies with $r>$ 1, no seeing correction is applied. Triangles and circles indicate uncorrected and corrected galaxies, respectively. Triangles can be interpreted as lower limit values. $\textit{Right:}$ Comparison between the observed and intrinsic angular momentum as a function of the galaxy S\'{e}rsic index for the corrected galaxies. Solid lines indicate the predicted behaviour of the corrections as a function of the S\'ersic index for the mean $r$ values of the respective subsets.}
         \label{fig:correc_lam}
    \end{figure*}
%%%%%%%%%%%%%%%%%%%%%%%%%%%%%%%%%%%%%%%%%%%%%%%%%%%%%%%

\subsection{Atmospheric seeing corrections for $\lambda_{R}$}\label{sec:seeing}

    The limited seeing due to turbulence in the atmosphere, the instrumental defects and the spatial sampling affect the measured values of the spin parameter by smearing the LOSVD due to the finite MUSE PSF. This results in the observed value of $\lambda_{R}$ being lower than the intrinsic one, and influences its radial distribution. At higher redshifts, both the apparent size of galaxies and the spatial resolution decrease, and consequently the seeing effect becomes more relevant. We explored three $\lambda_{R}$ seeing corrections. Firstly,   \cite{2018MNRAS.477.4711G} corrections derived by using realistic Jeans Anisotropic Models (JAM)  \citep{2008MNRAS.390...71C} of galaxy kinematics, which take into account the S\'{e}rsic index and the ratio of the seeing to the galaxy effective radius and can be applied to any IFS data set. Secondly,  \cite{2020MNRAS.497.2018H} corrections derived from a series of mock observations of N-body galaxy models, which compared to the latter also includes the effect of galaxy ellipticity as a proxy for the galaxy inclination. Lastly, \cite{2021MNRAS.505.3078V} corrections which are formally analogous to the \cite{2020MNRAS.497.2018H} corrections,  but are optimized for SAMI Galaxy Survey data. We selected the  \cite{2018MNRAS.477.4711G} method for two main reasons. Firstly, these corrections do not require the galaxy inclination as input, which is difficult to parametrize in observations without modelling. As our sample reaches low S/N galaxies, adding an extra parameter to the correction can increase the error in the corrected values of $\lambda_{R}$. In addition, the  \cite{2018MNRAS.477.4711G} corrections also allow us to correct the largest number of galaxies in the sample. 
    
    The function that describes the relation between the observed and the intrinsic spin parameter is given by
    \begin{equation}
        \lambda_{{\rm R}}^{{\rm obs}} =   \lambda_{{\rm R}}^{{\rm true}} gM_{2} \left ( \frac{\sigma_{{\rm PSF}}}{R^{\rm maj}} \right ) f_{n} \left ( \frac{\sigma_{{\rm PSF}}}{R^{\rm maj}} \right )
    \label{eq:correc}
    ,\end{equation} 
    where
    \begin{equation}
        gM_{2} \left ( \frac{\sigma_{{\rm PSF}}}{R^{{\rm maj}}} \right ) = \left [ 1 + \left ( \frac{\sigma_{{\rm PSF}} / R^{{\rm maj}}}{0.47}\right )^{1.76} \right ]^{-0.84}
    \label{eq:correc2}
    ,\end{equation} 
    and by
    \begin{equation}
        f_{n} \left ( \frac{\sigma_{{\rm PSF}}}{R^{{\rm maj}}} \right ) = \left [ 1 + (n-2) \left ( 0.26 \frac{\sigma_{{\rm PSF}}} {R^{{\rm maj}}}\right ) \right ]^{-1}
    \label{eq:correc3}
    ,\end{equation} 
    where $\lambda_{R}^{{\rm obs}}$ and $\lambda_{R}^{{\rm true}}$ are the observed (or measured) and the intrinsic (or true) angular momentum computed in a fixed aperture, respectively; $R^{{\rm maj}}$ is the semi-major axis of the aperture within which $\lambda_{R}$ is calculated; $n$ is the S\'{e}rsic index; and $\sigma_{{\rm PSF}}$ = ${\rm FWHM_{PSF}}/2.355$ (values provided by MUSE team, private communication); $f_{n}$ is an empirical function used to introduce the S\'{e}rsic index dependence into the correction; and $gM_{2}$ is the generalized form of the Moffat function \citep{1969A&A.....3..455M}. The correction is only appropriate for galaxies with an angular size larger than the PSF (FWHM)  (i.e. $\sigma_{{\rm PSF}} \leq R^{{\rm maj}}$).

    Figure \ref{fig:correc_lam} presents how the correction affects our galaxy sample. Of the 106 galaxies, 80 can be corrected. The correction is driven by the S\'{e}rsic index and by how well resolved the galaxy is, which is given by the parameter $r$ = $\sigma_{\rm{PSF}}/ \rm{R^{maj}}$. We   call $r$ the `resolution parameter' as it describes how well the aperture selected is to calculate $\lambda_{R}$ given the PSF. In the left panel, both corrected and non-corrected galaxies are presented. For the latter, the observed angular momentum can be treated as a lower limit. Non-corrected galaxies present a wide range of $r$ values, also smaller than 1. In their cases the correction is not applied due to unfavourable combinations of $r$ and $n$ values, which give a $\lambda_{R}^{{\rm true}}$ $>$ 1, and not because these galaxies have $r$ > 1. The region where the correction is unphysical is presented as a shaded grey area in Fig. \ref{fig:correc_lam}. Small galaxies, where the width of the Gaussian PSF is larger than their semi-major axis ($\sigma_{{\rm PSF}}>$ $R^{{\rm maj}}$ or $r$ $>$ 1) are also not corrected as they are simply too small.

        As mentioned previously, the correction depends on $r$ (or $\rm R^{maj}$ and $\sigma_{\rm{PSF}}$) and $n$. There are different values of $r_{max}$, defined as the maximum value of the resolution parameter $r$ for which the PSF correction can be applied to provide physically acceptable results (where $\lambda_{R}^{{\rm true}}$ $\leq$ 1), for different S\'{e}rsic indices (see Appendix \ref{sec:correc} for details). Both parameters, $r_{max}$ and $n$ are inversely related. Higher values of $n$, which means that the light distribution of a galaxy is more concentrated towards its centre, have lower $r_{max}$ values. This is what we observe in the left panel of Fig. \ref{fig:correc_lam}. There are possible combinations of $n$ and $r$ that push the corrected $\lambda_{R}$ into the non-physical region (grey shaded region), even if r < 1. We speculate that this is because our determination of these values ($\rm R^{maj}$, $n$, $\lambda_{R}$) is prone to error due to data quality.

    In the right panel of Fig. \ref{fig:correc_lam} we show the correction dependence on $n$ for the corrected galaxies in our sample. The S\'{e}rsic index for the corrected galaxies spans from $\sim$0.5 to $\sim$7.5. We can see that the correction  more strongly affects galaxies where their angular size is comparable to $\sigma_{{\rm PSF}}$; this is where $r$ approaches 1, as expected. The solid lines in the plot indicate how the generalized Moffat and $f_{n}$ (from Equations \ref{eq:correc2} and \ref{eq:correc3}) behave for different $n$ and $r$ values, which for the latter is the mean $r$ value for each subset. When the galaxy resolution decreases the correction is more affected by the galaxy light distribution (given by $n$), and therefore the ratio of the observed to intrinsic $\lambda_{R}$ increases. In other words, if the galaxy light is more concentrated towards its centre (higher $n$), a better resolution is required for the correction to be physically applicable. More details about the correction are presented in Appendix \ref{sec:correc}. 

\section{Results and discussion}\label{sec:results}

\subsection{Stellar kinematics maps}\label{sec:maps}

    We showed an example of the stellar kinematic maps in panels $(c)$ and $(d)$ of Fig. \ref{fig:example_gal}. The rest of our sample, including the resolved stellar kinematic maps and properties, are presented in Appendix \ref{sec:table_maps}. All the maps presented were constructed according to the criteria explained in Section \ref{sec:sample}. Categorizing these kinematic maps is not trivial, given the limitation of our data;  for several galaxies we do not reach a large number of bins due to the low overall S/N. As mentioned previously (Fig. \ref{fig:lam_dif_aper}), for most of the galaxies the data traces the kinematics up to 2$R_{e}$, so the spin parameter and the categorization of the maps for each galaxy were computed using the largest aperture possible. 

    Kinematic maps, in terms of morphology, allow  the galaxy sample to be separated into regular rotators (RRs) and non-regular rotators (NRRs), where the main distinction lies in how similar  the velocity map is to that of a stellar disc \citep{2008MNRAS.390...93K}. Galaxies were classified using \texttt{KINEMETRY} \citep{2006MNRAS.366..787K}, which performs a harmonic analysis of velocity maps, distinguishing between the regular (disc-like) rotation and complex velocity maps characterized by non-regular rotation. We built on the criteria for recognizing the  regular rotation established in \citet{2011MNRAS.414.2923K}, where RRs are defined as having $\overline{k_5/k_1}<0.04$, including the errors on the coefficients; $k_1$ and $k_5$ are harmonic coefficients referring to the rotational velocity (circular velocity in the case of pure discs) and the higher-order harmonic terms describing the deviations from the regular rotation, respectively. In addition to this criterion, we also added that the extent of the map analysed by kinemetry has to be at least 30\% larger than the size of the PSF, as measured by $\sigma_{\rm PSF}$, and the average value of $k_5/k_1$ is calculated for $r>\sigma_{\rm PSF}$. Galaxies with $\overline{k_5/k_1}>0.04$ beyond $\sigma_{\rm PSF}$ are classified as NRRs. Furthermore, galaxies for which we were not able to measure $k_5/k_1$ beyond $\sigma_{\rm PSF}$, or when the uncertainties on $k_5/k_1$ were too large for robust estimation, were classified as uncertain, where we distinguish two possibilities of RR$^*$ and NRR$^*$. We list in Table \ref{tab:sample_table} the classification of the velocity maps.
%%%%%%%%%%%%%%%%%%%%%%%%%%%%%%%%%%%%%%%%%%%%%%%%%%%%%%%
    \begin{figure}
    \centering
       \includegraphics[width=9.0cm]{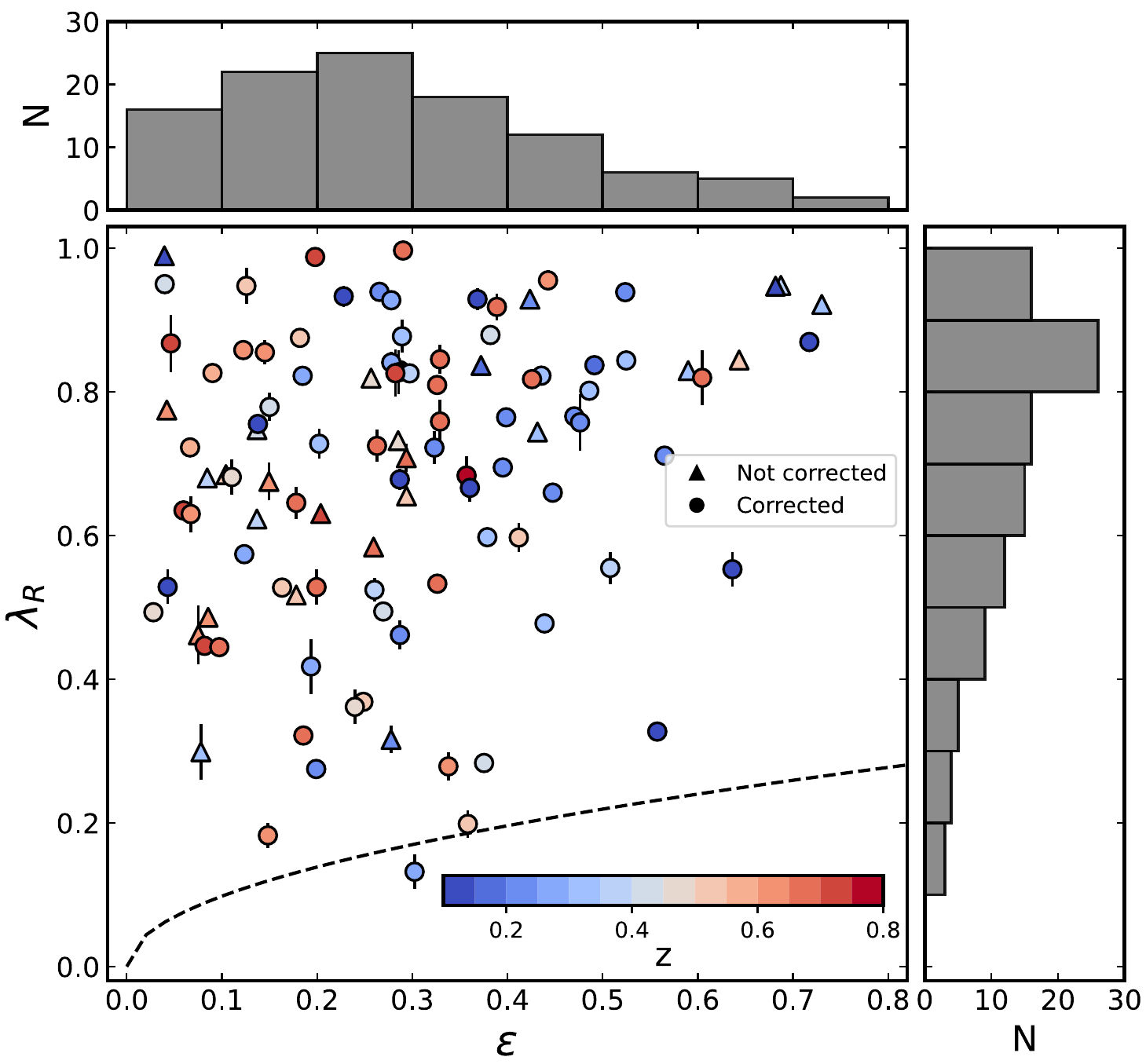}
         \caption{$\lambda_{R}$ as a function of the ellipticity coloured by the redshift for the 106 galaxies in the sample, with the respective histograms. The circles  indicate galaxies where the spin parameter was corrected by the atmospheric seeing, and therefore the value corresponds to the intrinsic stellar angular momentum ($\lambda^{{\rm true}}_{R}$). The triangles are galaxies where corrections  could not be applied, and thus the values are the measured ones ($\lambda^{{\rm obs}}_{R}$) and can be considered  lower limits. The black dashed line denotes the threshold between slow and fast rotators, {given by 0.31 $\times \sqrt{\epsilon}$ \citep{2011MNRAS.414..888E}}.}
         \label{fig:lam_e}
    \end{figure}
    
%%%%%%%%%%%%%%%%%%%%%%%%%%%%%%%%%%%%%%%%%%%%%%%%%%%%%%%
    \begin{figure*}
    \sidecaption
       \includegraphics[width=12cm]{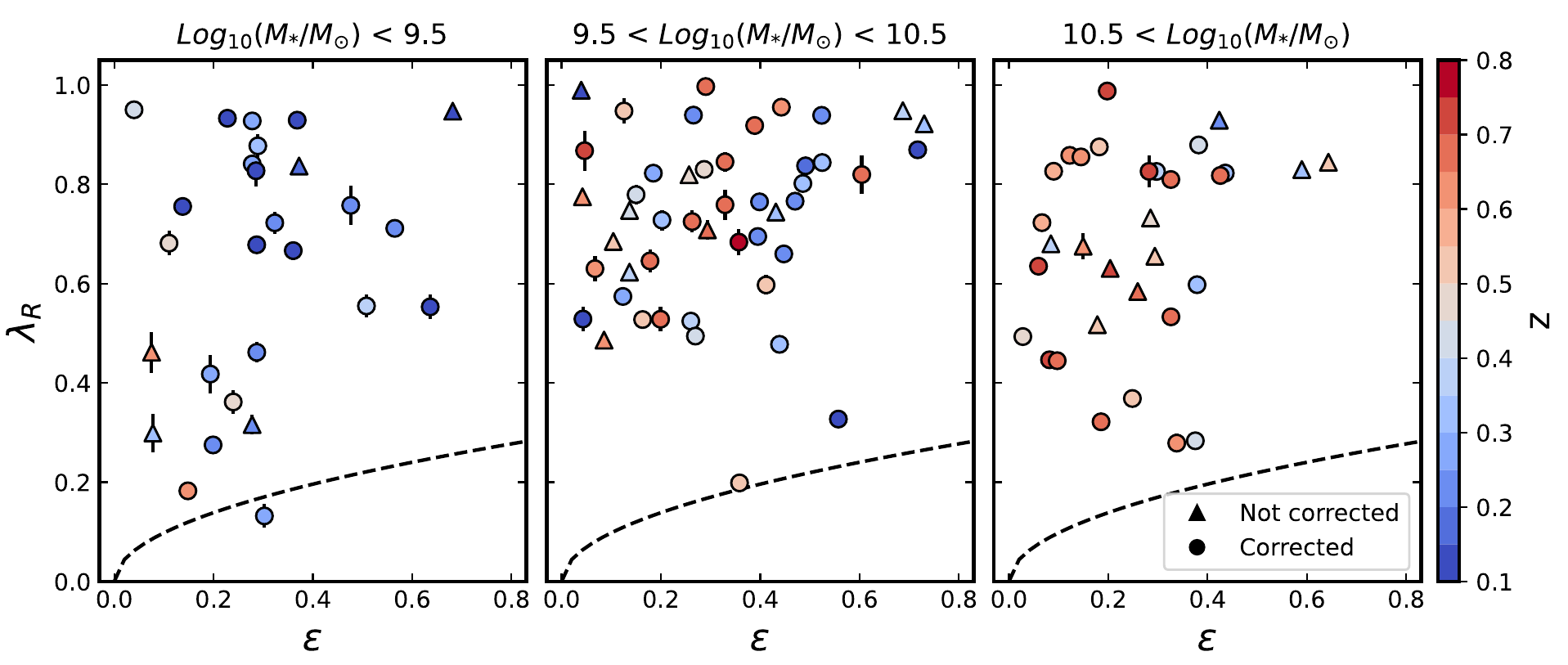}
         \caption{$\lambda_{R}$ as a function of the galaxy ellipticity coloured by the redshift for three different mass bins indicated in the top part of each panel. The black dashed lines denote the threshold between slow and fast rotators. Uncorrected and corrected $\lambda_{R}$ values are indicated by triangles and circles, respectively.}
         \label{fig:e_lam_mass}
    \end{figure*}
%%%%%%%%%%%%%%%%%%%%%%%%%%%%%%%%%%%%%%%%%%%%%%%%%%%%%%%

\subsubsection{Regular rotators}\label{sec:RR}

    A significant fraction of galaxies in our sample ($\sim$40\%) exhibit clear stellar rotation with velocity maps characteristic of RRs and with typical amplitude values ranging to $\pm$100 km s$^{-1}$ and in some cases to $\pm$200 km s$^{-1}$. Another $\sim$30\% of galaxies are classified as RR$^*$, indicating that the majority of the galaxies in the sample are expected to be RRs. Most RRs are disc-like galaxies, where their flattened nature is visible; for example galaxies 4, 12, 17, 21, 34, 68, 85, and 88 are edge-on galaxies, some potentially spirals, or flat ETGs with clear rotation in their stellar velocity fields and generally constant velocity dispersion maps. Various galaxies present clearly visible spiral arms as well, for example galaxies 7, 16, 53, 84, 86, and 97 have face-on spiral morphology, while some also have obvious bars, for example  galaxies 57, 76, 81, 86, and 92, implying that they are all disc-dominated systems. 

    There are also sample galaxies with smooth and roundish light distribution without identifiable spiral structures. Some cases are galaxies 25, 26, 30, 45, 48, 55, 56, 94, and 99. These galaxies also look compact in HST images. Most of them have constant stellar velocity dispersion maps, except for galaxies 45 and 56, which  present a peak at their centres. Galaxies  25, 26, 30, and 56 present very low levels of star formation, allegedly quenched. As galaxies at higher redshift appear rounder (see Fig. \ref{fig:lam_e}), they could be more affected by the instrumental PSF due to the decrease in angular size with redshift. 

    The majority of the velocity dispersion fields are constant across the map except galaxies 32, 53, 57, 98, and 101, which show a central peak.  Some galaxies have lenticular shapes easily recognized in, for example, these edge-on cases: galaxies 15, 34, 43, 68, 88, 85, 96, and 101. A few galaxies have velocity dispersion maps with peaks off centre, which are likely explained by large uncertainties (e.g. galaxies 29, 34, and 85), or are caused by possible interactions (e.g. galaxy 27). Galaxy 76 shows a stellar velocity map with symmetry axis along the major axis, and while the HST image reveals spiral and bar structures, the velocity dispersion map does not present particular features. 
    The difference between the kinematic and photometric axis is a consequence of the kinematics tracing the motion of stars in the disc, as seen in a number of other bar systems \citep[e.g.][]{2011MNRAS.414.2923K,2020A&A...643A..14G}.
    %ID 87 could have a weak disc but has a clear ordered rotation and a central peak velocity in its dispersion map.

    Some interesting RR pairs are the galaxies  1--2 and 69--70. They could be interacting given their angular proximity and similar redshifts. Galaxies  1 and 2 have similar masses of the order of $\sim$10$^{10.6}$ M$_{\odot}$, and have irregular morphology at large radii, possibly from the interaction, but both present nuclear bulges. For galaxy 1 it is possible to see a distorted spiral or a tidal arm, and for galaxy 2 a distorted disc, given the inclination of the galaxy. The  stellar velocity dispersion map of galaxy 1 is more centrally peaked compared to galaxy 2, which is overall flat. For galaxies 69 and 70 the  shape of the individual galaxies does not seem to be distorted. Galaxy 69 has a stellar mass of $\sim$10$^{9.7}$ M$_{\odot}$, with a spiral morphology; a disc and blurry arms can be seen in the HST images. Galaxy 70, which is more massive ($\sim$10$^{10.3}$ M$_{\odot}$), has a global spectrum typical of an early-type galaxy, and the imaging does not show any spiral structure but a smooth light distribution concentrated in its centre, with a fitted S\'{e}rsic index ($n$=4.7) similar to a de Vaucouleurs profile. The velocity dispersion maps of both galaxies (69 and 70) show centrally peaked stellar velocity dispersion maps. 

In both pairs (1--2 and 69--70) the measured angular momentum for the interacting galaxies is comparable $\lambda_{R} \sim$0.7, and the velocity maps look regular so the rotation of the galaxies seems not to be affected by the closeness. However, they are too small for robust determination, and they are both classified as RR$^*$.

\subsubsection{Non-regular rotators}\label{sec:NRR}

    There are also NRRs in the sample, but their fraction is significantly lower compared to the RRs, about 22\%, and we classified around 7\% of galaxies as NRR$^*$. This is potentially influenced by the difficulties in classifying certain galaxies, due to their size, but only a small fraction of RR$^*$ galaxies are expected to be misclassified as NRR$^*$. Some NRR examples are galaxies 6, 23, 38, 77, 93, and 103, where the stellar velocity maps present rotational fields that are more complex, frequently because the total number of bins is small and their spatial size are large. This limits the information for the kinematic analysis, but also the size of the bin in mixing regions within the galaxy that can have different kinematics; therefore, the outcome is a complex velocity map. Morphologically, galaxy 5 has an irregular appearance with a displaced brighter central region which is where the maximum velocity dispersion is found in the $\sigma_{\rm{stellar}}$ map. Galaxy 13 presents no morphological features and a round shape with highly concentrated light distribution, while its velocity map is irregular and characterized by low velocities. Galaxy 20 has a centrally peaked stellar velocity dispersion map, but  the HST images show two objects that can be recognized, which are blended in the MUSE data. The blending is likely the reason for the asymmetry in the velocity map. For galaxies 38 and 77, no visible structures are seen; they have flat velocity dispersion maps and a relatively high number of bins. Galaxy 40 is an example of a possible prolate rotator, with rotation around the minor axis, but it is too small to robustly establish this fact, including a possible central counter rotation.  Galaxies 90, 93, and 95 are round systems and have small angular sizes, and therefore are strongly affected by the PSF. 
    
    One special case is  galaxy  10. The velocity map shows counter-rotation in the centre with respect to the outskirts. The $\sigma$-map does not have a central peak, but there are indications of high-velocity dispersion on both sides of the centre. In both respects, this galaxy has characteristics of a 2$\sigma$ galaxy \citep{2011MNRAS.414.2923K}, which is interpreted as having two counter-rotating discs, a combination of which makes the observed rotation and $\sigma$ maps. Regarding the HST image, the galaxy has a lenticular and featureless light shape.
 %%%%%%%%%%%%%%%%%%%%%%%%%%%%%%%%%%%%%%%%%%%%%%%%%%%%%%%
    \begin{figure*}
    \centering
       \includegraphics[width=18cm]{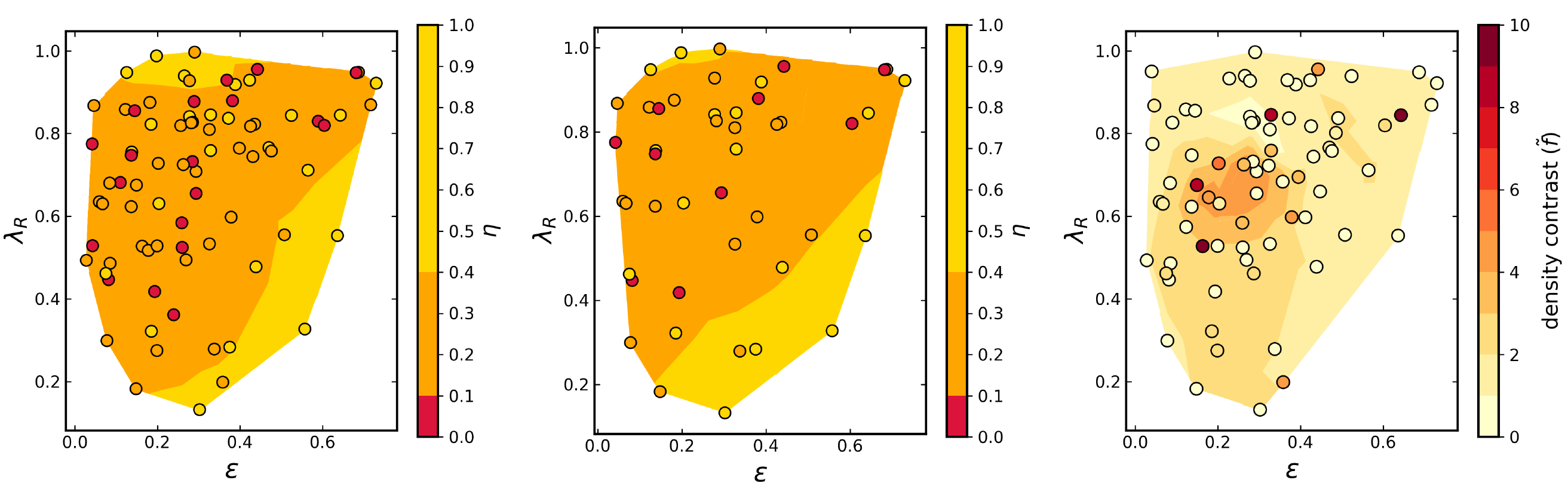}
         \caption{$\lambda_{R}$-$\epsilon$ plane coloured by two different environmental indicators. The background for all panels is a LOESS-smoothed version of the dots above. \textit{Left:} $\eta$ indicator computed from the use of the FoF algorithm for all galaxies detected in groups. \textit{Middle:} $\eta$ indicator computed from the use of the FoF algorithm for galaxies detected in groups with $\geq$ 10 members. \textit{Right:} Density contrast ($\tilde{f}$) estimator obtained from the Voronoi tessellation method.}
         \label{fig:env}
    \end{figure*}
%%%%%%%%%%%%%%%%%%%%%%%%%%%%%%%%%%%%%%%%%%%%%%%%%%%%%%%   
\subsection{Stellar spin parameter}

    We compute the galaxy spin parameter $\lambda_{R}$, as mentioned in Section \ref{sec:lam}, using the information presented in the stellar kinematic maps for different apertures between 0.5 and 2$R_{e}$. Most of the sample galaxies have flat $\lambda_{R}$ profiles (see Fig. \ref{fig:lam_dif_aper}), so the aperture chosen as a final spin parameter measurement for each system is the largest one possible. Selecting the largest aperture allows us to reduce the impact of the seeing effect in the measured stellar angular momentum and to correct the observed $\lambda_{R}$ for the largest number of galaxies in the sample (see Section \ref{sec:seeing}). The details of the aperture chosen for every galaxy are given in Table \ref{tab:sample_table}. 
    
    Figure \ref{fig:lam_e} shows the stellar angular momentum $\lambda_{R}$ as a function of the observed ellipticity $\epsilon$, coloured by the spectroscopic redshift for all galaxies in the sample. Circles indicate spin parameter values corrected by the instrumental PSF and triangles non-corrected observed values of $\lambda_{R}$ and can be considered as lower limits. The black dashed line denotes the separation between slow and fast rotators, given by 0.31 $\times \sqrt{\epsilon}$ \citep{2011MNRAS.414..888E}. 

We can see several aspects regarding this diagram. Firstly, the sample is dominated by galaxies with high values of spin parameter, which is consistent with the high fraction of kinematic maps classified as regular rotators (RR) (see Section \ref{sec:maps}), and the distribution of the spin parameter is globally increasing for flatter galaxies and there are only a few flat galaxies with low stellar angular momentum.
    Secondly, there are basically no slow rotators in the sample, with the exception of one galaxy (73) where our statement holds at a 3$\sigma$ level. Galaxy  73 is at z = 0.28, has a stellar mass of  10$^{8.4}$ M$_{\odot}$, a star formation rate of log$_{10}$(SFR) = -0.7 M$_{\odot}$yr$^{-1}$, and is therefore  located on the star-forming main sequence. Even if the value of $\lambda_{R}$ is seeing-corrected for this galaxy, the system is dominated by the PSF due to its small angular size, and it is still possible that we measure only the upper limit to its stellar angular momentum. This galaxy has inherently disturbed kinematics, as suggested from the kinemetric analysis (see Section \ref{sec:maps}). This means that galaxy 73 falls in the group of low-mass galaxies where the angular momentum is lowered due to gas-rich interactions, either through accretion of counter-rotating gas, which subsequently forms a counter-rotating stellar components, or a more general gas-rich merger, which perturbs the observed kinematics \citep{2020A&A...635A.129K}. While the final stellar kinematics of this galaxy are still uncertain, we can conclude that there are no massive slow rotators in our sample. 
    Thirdly, the majority of our sample galaxies tend to have smaller values of ellipticity ($\epsilon <$ 0.4), and thus their shapes tend to be round. Galaxies with the highest ellipticity values ($\epsilon \gtrsim$ 0.7;  4, 21, 34, and 68) have low redshifts z $\approx$ 0.1-0.3 and are classified as edge-on spirals. In Fig. \ref{fig:e_lam_mass} we present the same diagram as in Fig. \ref{fig:lam_e}, but binned in three different mass bins. Galaxies at low redshift have predominantly stellar mass $\leq$ 10$^{9.5}$ M$_{\odot}$, present a wide variety of ellipticites and $\lambda_{R}$ as well. When the stellar mass increases, the fraction of round galaxies at higher redshift also increases. Galaxies in the mass bin $\geq$10$^{10.5}$ M$_{\odot}$ are primarily round galaxies at z > 0.5.

    In the local Universe, slow rotators are massive, round galaxies, and dominate above $M_{crit}\sim$ 10$^{11}$ M$_{\odot}$ \citep{2016ARA&A..54..597C}. In our sample, only six galaxies are more massive than the critical mass (23, 44, 48, 51, 77, and 79) and are at z > 0.4. They show both regular and non-regular rotation in their velocity maps. The absence of slow rotators therefore might not be so surprising as we do not have analogues to nearby galaxies.

\subsection{Environment indicators}

    We would like to explore if there is any dependence between the stellar angular momentum of the galaxies and their environment. Therefore, we first quantify the environment for the galaxy sample. There are different methods for identifying structures and obtaining density estimators. For our work, we  explore two separate approaches. Based on the spectroscopic data for MUSE fields, we identify galaxy structures using the friends-of-friends (FoF) algorithm \citep{1985ApJ...292..371D}. The FoF algorithm searches friends in a volume that is dependent on the angular and redshift separation. We use a  method similar to that described in  \citet{2023arXiv231200924E}. After groups are identified, their properties, such as $R_{200}$ (the radius where the density of the group is equivalent to 200 times the Universe's critical density), virial mass, and radius, can be derived. The FoF algorithm was run in our sample identifying 93 of the total 106 galaxies belonging to a group with $\geq$ 2 members. 
    
    For galaxies in structures, a global density estimator can be calculated, which is given by the dimensionless parameter $\eta$ \citep{2013MNRAS.436L..40N}: 
    \\
    \begin{equation}
        \eta = \frac{\mid \Delta v \mid}{\sigma_{g}} \frac{\Delta r}{R_{200}}
    .\end{equation} 
    \\
    Here $\Delta v$, $\Delta r$, and $\sigma_{g}$ are the velocity of the galaxy within the group, the projected distance to the group centre, and the velocity dispersion of group members, respectively, and  $\eta$ is a measurement for how dynamically tied a galaxy is to its group. Galaxies with $\eta$ < 0.1 are considered early accreted by their host. If they already passed the pericentre of their orbit once 0.1 < $\eta$ < 0.4 is expected, and if they have recently been accreted by the group 0.4 < $\eta$ < 2 is anticipated \citep{2013ApJ...768..118N}. The parameter is reliable for structures with well-defined properties, which is in general when the group richness or the number of galaxy members is equal to or higher than 10 \citep{2023arXiv231200924E}. Around 45$\%$ of all galaxies within a group in our sample have $\geq$10 members. 

    In the left and middle panels of Fig. \ref{fig:env} we present the $\lambda_{R}$-$\epsilon$ plane coloured by $\eta$ for all sample galaxies in groups and for galaxies with $\geq$10 members, respectively. As we have a sparse data set of a few objects only, we smoothed them using the locally weighted regression method called locally estimated scatterplot smoothing (LOESS) \citep{Cleveland1988LocallyWR} implemented by \cite{2013MNRAS.432.1862C}, which is shown in the background of both panels. We do not see a significant correlation between the level at which the galaxy is bound to its group and the spin parameter. We do see, however, that galaxies with low $\eta$ ($\eta$ < 0.1) present a wide variety of spin parameter values. In addition, galaxies that were early accreted by their groups, where the group has $\geq$10 members, have higher stellar angular momentum ($\lambda_{R} \gtrsim$ 0.4).

    We also quantify the environment for our sample using another parameter. Utilizing Voronoi tessellation \citep{2001A&A...368..776R} we can estimate the surface density of galaxies by measuring the environment overdensities. The method allows us to  divide a distribution of points on a plane into convex cells or polygons, where every cell only contains a unique point as the bin generator. The bin generator has the property that its cell vertices are nearer to it than to any other point on the plane. We compute a local density estimator given by $\tilde{f}$, named local density contrast \citep{2021ApJ...911...46S}, defined as ($\tilde{f}=f/\langle f \rangle$). The Voronoi tesselation has the mathematical property that the local density $\tilde{f}$ corresponds to the inverse of the cell area ($a$) where ($\langle f \rangle = \langle 1/a \rangle$) is the mean density of all cells. Small bins that are equal to small-area polygons are denser regions, and large bins correspond to low-density regions, in relative terms. 

    The density estimator was computed by building redshift slices of $\pm$3.7 Mpc thick around each galaxy \citep{2023arXiv231200924E}, and therefore the physical volume for each galaxy was similar. Around 15$\%$ of the galaxies are impacted by edge effects due to MUSE surveys footprint. In the right panel of Fig. \ref{fig:env}, we present the density contrast compared to the stellar spin parameter and galaxy ellipticity. There is a tendency for galaxies in denser environments (higher density contrast) to have $\lambda_{R} \gtrsim$ 0.5 and to present a wide range of ellipticity values.
    
    The methods are completely different mathematically; the two methods target different environment scales: the FoF algorithm identifies galaxy structures, and therefore explore several megaparsecs. On the other hand, Voronoi tessellation probes smaller scales (no more than $\sim$4 Mpc). Comparing both density estimators, although radically different, provides consistent results that the environment does not have a strong influence on $\lambda_{R}$, but also that we are not probing very dense regions. 

\subsection{Stellar angular momemtum and comparison with different surveys}\label{sec:maps_slow}

    We would like to compare our results with different local surveys that have in the past measured the stellar angular momentum for galaxies. However, these works are not directly comparable for several reasons, first because of the sample selection and galaxy properties, for example, the galaxy mass range. For the majority of z=0 surveys (e.g. SAMI, \citealp{2012MNRAS.421..872C}; MASSIVE, \citealp{2014ApJ...795..158M}, ATLAS$^{3D}$ \citealp{2011MNRAS.413..813C}), the environment is well characterized or it was taken into account when the sample was constructed. SAMI is a stellar-mass-selected survey containing both early- and late-type galaxies. MASSIVE is a stellar mass selected survey that targeted massive galaxies with masses M$_{\star}\gtrsim$10$^{11.5}$ M$_{\odot}$ in diverse galaxy environments. ATLAS$^{3D}$ was built by selecting nearby ETGs. In our case we did not apply any mass or environment cut; we included all the galaxies with enough S/N in the redshift range available where kinematic maps could be built. Secondly, our sample covers a large redshift range and is actually very small at any given redshift bin to study the evolution of the spin parameter at similar cosmic ages. In addition, the lack of slow rotators in our sample makes us unable to track when or how the slow rotator population started to build up. The deficit of slow rotators could be due to our objects typically sampling the field population of galaxies and, more importantly, a high fraction of galaxies in the sample present significant star formation levels, while slow rotators are expected to be massive quiescent galaxies \citep[e.g.][]{2020MNRAS.495.1958W,2021MNRAS.503.4992F}. While around 30$\%$ of the sample galaxies are quiescent or with low levels of star formation, only about 5\% are galaxies with stellar masses higher than $10^{11}$ M$_\odot$ (see Fig. \ref{fig:red_mass_sfr}).

\begin{figure}
    \centering
       \includegraphics[width=9.1cm]{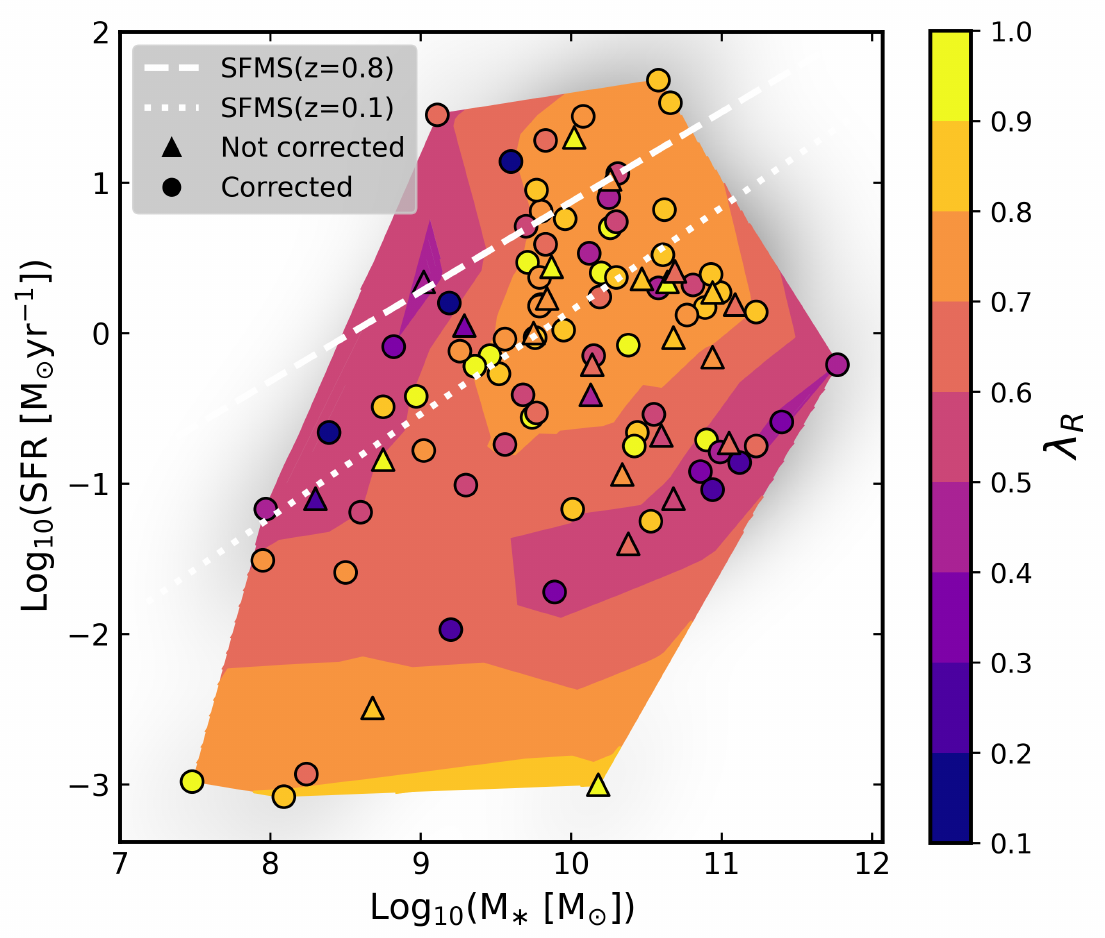}
         \caption{Star formation rate  as a function of the stellar mass coloured by $\lambda_{R}$. The grey-shaded region in the background corresponds to the 3D-HST sample measurement. The dotted and dashed white lines indicate the star formation main sequence  for an age of the Universe of 12.4 Gyr (z=0.1) and 6.8 Gyr (z=0.8) given by Eq. 1 of \cite{2012ApJ...754L..29W}, respectively. The colour map is a LOESS-smoothed version of the markers on top. Dots and triangles are seeing and not seeing corrected values, respectively.}
         \label{fig:e_lam_sfr}
    \end{figure}

    Nevertheless, we can pursue this comparison further, and in Fig. \ref{fig:e_lam_sfr} we perform a similar analysis to that of \cite{2020MNRAS.495.1958W}, showing the SFR - M$_{\star}$ plane colour-coded by $\lambda_{R}$. We smooth the $\lambda_{R} - \epsilon$ diagram using the LOESS method, which results in the coloured background.  We see that the majority of galaxies with spin parameters > 0.7 are in or around the so-called star-forming main sequence, which is consistent with the scenario where star-forming galaxies have disc-like structures, and therefore, high levels of rotation. There  are also galaxies with low levels of star formation, but high $\lambda_{R}$ (39, 46, 60, 97, and 98), similar to local ETG fast rotators. Some quiescent galaxies have complex stellar kinematic maps, small angular size, or some  unresolved structures that can be seen in the HST image but not in MUSE data. Thus, it is hard to characterize the spin parameter behaviour for galaxies with log$_{10}$(SFR)$\leq$ -2 M$_{\odot}$yr$^{-1}$ in our sample. Even with all caveats associated with our data (small sample, few galaxies at any given redshift, low S/N and small sizes), we find a similar result to those of  \cite{2020MNRAS.495.1958W}: there is a hint that quiescent galaxies or galaxies with low levels of star formation   have lower values of angular momentum.

    We also find galaxies on the main sequence of star formation with lower values of the spin parameter. These galaxies are known to exist in the local universe  \citep[][see their Fig. 1]{2020MNRAS.495.1958W}, and are also consistent with the present ETG population, which contains low-mass ($<10^{11}$M$_\odot$) slow rotators \citep{2011MNRAS.414..888E}. The stellar angular momentum in these galaxies is lowered due to interactions (i.e. gas accretion or ongoing merger) that perturb the stellar disc \citep{2020A&A...635A.129K}. In some cases the final outcome will be galaxies with artificially lowered angular momentum due to counter-rotating components, but it is also possible that the stellar disc will reform and the galaxy will end up as a fast rotator. Formally, our only slow rotator (galaxy 73) is such a galaxy, but it is too small for a robust determination of its kinematic state. Another galaxy (10) is a fast rotator, but it shows evidence of disc counter-rotation, and is one of the rare cases of flat galaxies with low $\lambda_{R}$ that do not have their velocity anisotropy related to their (intrinsic) flattening, as is expected for the general population of fast rotators \citep{2007MNRAS.379..418C}.
    
    {As stated above, our sample is limited by the signal-to-noise ratio when extracting the stellar kinematics of the galaxies. The S/N for our sample is influenced by the galaxy brightness and also by the angular size, given that the galaxies have to be kinematically resolved. Therefore, as a consequence, the sample has an implicit bias in galaxy mass and size. At higher redshifts (z > 0.5) we lack small galaxies with stellar mass M$_{\star}$< 10$^{9}$ M$_{\odot}$ (see left panel of Fig. \ref{fig:red_mass_sfr}), which could grow in both, mass and size, and be sampled by z = 0 surveys. However, we expect that the majority of these galaxies will kinematically evolve to become fast rotators, as they will most likely grow by accretion. As we found that $\sim$99\% of the galaxies in our sample are fast rotators, these low-mass-at-high-redshift galaxies will not contribute to a \textit{progenitor bias} for our case.}

    {The classical \textit{progenitor bias} \citep{2001ASPC..230..581F} is defined in terms of early-type galaxies. If we make an analogy, but instead of ETGs we follow the evolution of slow rotators, we can define a `kinematic' progenitor bias. This bias results from selecting two samples of slow rotators, one at z $\sim$ 0 and the other at high redshift; the sample of high-redshift slow rotators will underestimate the fraction of low-redshift slow rotators, given that also high-redshift fast rotators could evolve to be part of the slow-rotating population at z = 0.}

    {Nevertheless, in our sample this potential bias is different. In the local Universe, around 10$\%$ of the galaxy population with M$_{\star}$ > 10$^{9}$ M$_{\odot}$ are slow rotators \citep{2018MNRAS.477.4711G}. In our redshift range, we only find fast rotators. Hence, the kinematic progenitor bias in our sample arises from the non-existence of slow rotators at z > 0.5. If indeed there are no slow rotators at these redshifts, then a strong evolution up to z = 0 is expected. On the contrary, if there exists a population of slow rotators at z > 0.5 that our sample misses, there are two possible conclusions. Firstly, our sample is biased towards no slow rotators because we are not probing dense regions or massive galaxy structures, where slow rotators in the nearby Universe are found. Secondly, our sample is missing very massive galaxies (M$_{\star}$ > 10$^{11.5}$ M$_{\odot}$) at each redshift range, simply because they are missing from the footprint of the MUSE surveys.}
 %%%%%%%%%%%%%%%%%%%%%%%%%%%%%%%%%%%%%%%%%%%%%%%%%%%%%%%
%%%%%%%%%%%%%%%%%%%%%%%%%%%%%%%%%%%%%%%%%%%%%%%%%%%%%%%

\section{Conclusions}

    We measured the stellar angular momentum parametrized by the spin parameter $\lambda_{R}$ from the stellar kinematics of 106 galaxies in various MUSE surveys. The sample contains galaxies in the redshift range of 0.1 < z < 0.8, stellar masses from $\sim$10$^{7.5}$M$_{\odot}$ to 10$^{11.8} $ M$_{\odot}$, and  with star formation rates from log$_{10}$(SFR) $\approx$-3 M$_{\odot}$yr$^{-1}$ to $\approx$1.7 M$_{\odot}$yr$^{-1}$. The spin parameter was computed within 0.5-2 $R_{e}$ for each galaxy. As the galaxy kinematics are affected by the atmospheric seeing, we corrected the observed $\lambda_{R}$ by the instrumental PSF. Of the total 106 galaxies, 80 can be corrected. 

    We discussed in detail the stellar velocity and velocity dispersion maps of the sample, relating them to the galaxy morphology and classifying them into regular and non-regular rotators. We found that the significant fraction of galaxies in the sample ($\sim$40$\%$) present stellar velocity fields characteristic of RRs with a high fraction of nearly constant stellar velocity dispersion maps, which is consistent with the presence of discs or disc-like structures. In addition, the NRRs in the sample present more complex rotational fields, showing, in general, rotation without a clear symmetry axis. Some kinematic maps are not easily classified due to the combination of the low S/N, small sizes, and instrumental resolution. 

    Through the analysis of the $\lambda_{R}$ - $\epsilon$ diagram, we see no evident correlation or evolution between the spin parameter and redshift. Most of the sample is dominated by round galaxies, which tend to be found at higher redshifts and are also  the most massive galaxies. The spin parameter distribution increases towards one, peaking at $\lambda_{R} \sim$ 0.8. This is consequent with the fraction of regular rotators found in the sample. We do not find any significant population of slow rotators, but detect one galaxy consistent with being a low-mass slow rotator. We also see that massive quiescent or low-level star-forming galaxies present in general low values of $\lambda_{R}$ in the sample, which can support the scenario where quenching mechanisms are responsible for the lowering of the angular momentum in galaxies. Some star-forming galaxies also have low $\lambda_{R}$, which is a consequence of their perturbed kinematics, due to possible gas-rich interactions.

    Furthermore, we characterize the environment for our galaxy sample using two different indicators. Neither the global density estimator ($\eta$) for galaxies in structures nor the local density contrast ($\tilde{f}$), which measures overdensities, follows an evident trend with the stellar angular momentum. Thus, the environment at small and large scales does not seem to have any effect on the stellar angular momentum in our sample. This may mean that the rotation in our sample galaxies is mostly driven by internal processes rather than being influenced by the low to moderately dense environment. We note that the studied galaxies are not found in dense environments. It will be   necessary to probe these environments to establish any possible relation with $\lambda_{R}$ at z > 0. 
    
\begin{acknowledgements}
      This work was supported by the German
      \emph{Deut\-sche For\-schungs\-ge\-mein\-schaft, DFG\/} project
      number 4548/4-1. T. Urrutia acknowledges funding from the \emph{ERC-AdG grant SPECMAP-CGM-101020943}. L. A. Boogaard acknowledges support by \emph{ERC AdG grant 740246 (Cosmic-Gas)}. 
\end{acknowledgements}

\clearpage
\onecolumn

\appendix
\section{Seeing corrections \citep{2018MNRAS.477.4711G}} \label{sec:correc}

The correction by \cite{2018MNRAS.477.4711G} is derived from JAM models with full coverage within the effective radius and relies on the assumption that the measured $\lambda_{R}^{\rm{obs}}$ profile follows Equation \ref{eq:correc}. The correction is accurate for galaxies with 0.5 $\leq n \leq$ 6.5 and regular rotators (spirals and early-type galaxies). As non-regular rotators have inherently low $\lambda_{R}\leq$ 0.2 the expected seeing effect is small. 

\begin{figure}
    \centering
        \includegraphics[width=8.5cm]{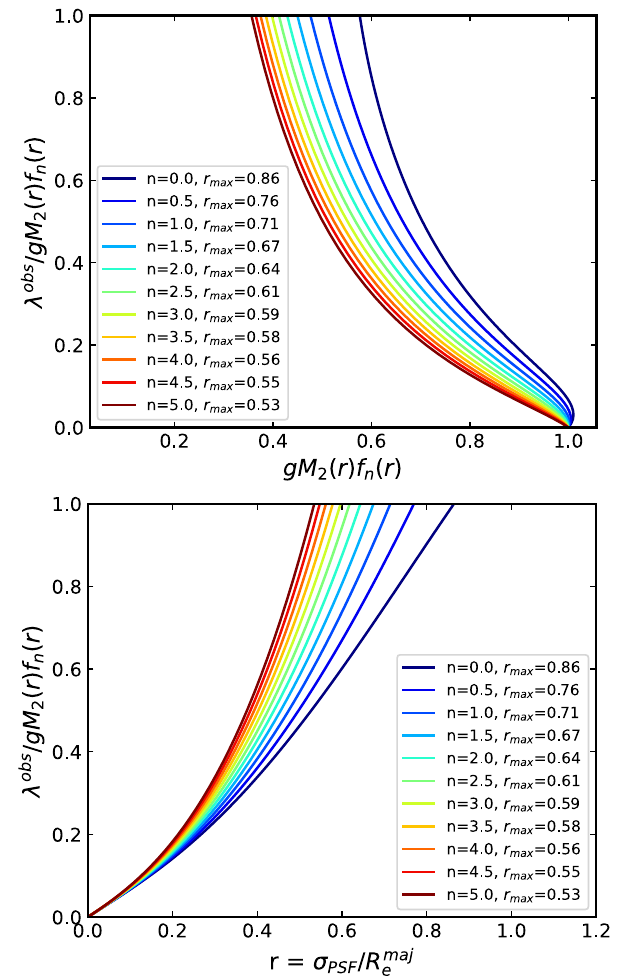}
         \caption{Behaviour of \cite{2018MNRAS.477.4711G} correction: y-axis in both panels is a proxy for the \textit{true} or corrected value of the angular momentum, which is dependent on galaxy resolution (given by $r$), the observed angular momentum ($\lambda^{obs}$), the generalized Moffat (gM$_{2}$), and the empirical $f_{n}$ functions. The upper panel shows the true $\lambda$ as a function of the functions that model the correction. Different curves show different S\'{e}rsic index values and the minimum galaxy resolution ($r_{max}$) where the correction is in the physical space ($\lambda^{true}\leq$1). The lower panel shows the direct dependence of the  true $\lambda$ with the resolution parameter $r$ values.}
         \label{fig:correc_an}
    \end{figure}

Here we explore how the generalized Moffat function (gM$_{2}$(r)) and the empiral function $f_{n}$ behave for different values of  resolution parameter ($r$ = $\sigma_{\rm{PSF}}/ \rm{R^{maj}}$) and S\'{e}rsic index ($n$). In Fig. \ref{fig:correc_an} we see the limits where the correction can be applied for any value of the observed angular momentum, these limits depend on $n$ and on how well resolved  the galaxy is. Higher $r$ values mean that the spatial resolution is low; in other words, that  the galaxy effective radius is similar to the width of the PSF ($r\sim$1). For example, for a physical correction of a galaxy with $n$ = 3, the resolution parameter has to be $r <$0.59, as shown by the green line in Fig. \ref{fig:correc_an}. The lower panel contains information on how the PSF and aperture sizes affect the true values of the spin parameter, while the upper panel shows the effects on the functions that model the correction. 

We conclude that for every value of $n$ there exists a maximum $r_{max}$ where the correction gives $\lambda_{R}^{\rm{true}}$ =  $\lambda_{R}^{\rm{obs}}$/gM$_{2}$(r)$f_{n}$ lower than 1. The relation between $n$ and $r_{max}$ is inverse: the higher the surface brightness is in the galaxy centre or the steeper the profile is, the lower $r_{max}$ is or equivalently the ratio of the width of the PSF and the galaxy size has to decrease. Therefore, the resolution parameter has to improve in order to  apply the seeing correction. In other words, if the galaxy light is more concentrated towards its centre (higher $n$), the resolution achieved has to improve for the correction to be physically applicable. 

%%%%%%%%%%%%%%%%%%%%%%%%%%%%%%%%%%%%%%%%%%%%%%%%%%%%%%%
%%%%%%%%%%%%%%%%%%%%%%%%%%%%%%%%%%%%%%%%%%%%%%%%%%%%%%%

\section{Galaxy sample}\label{sec:table_maps}

\begin{figure*}
    \centering
       \includegraphics[width=19cm]{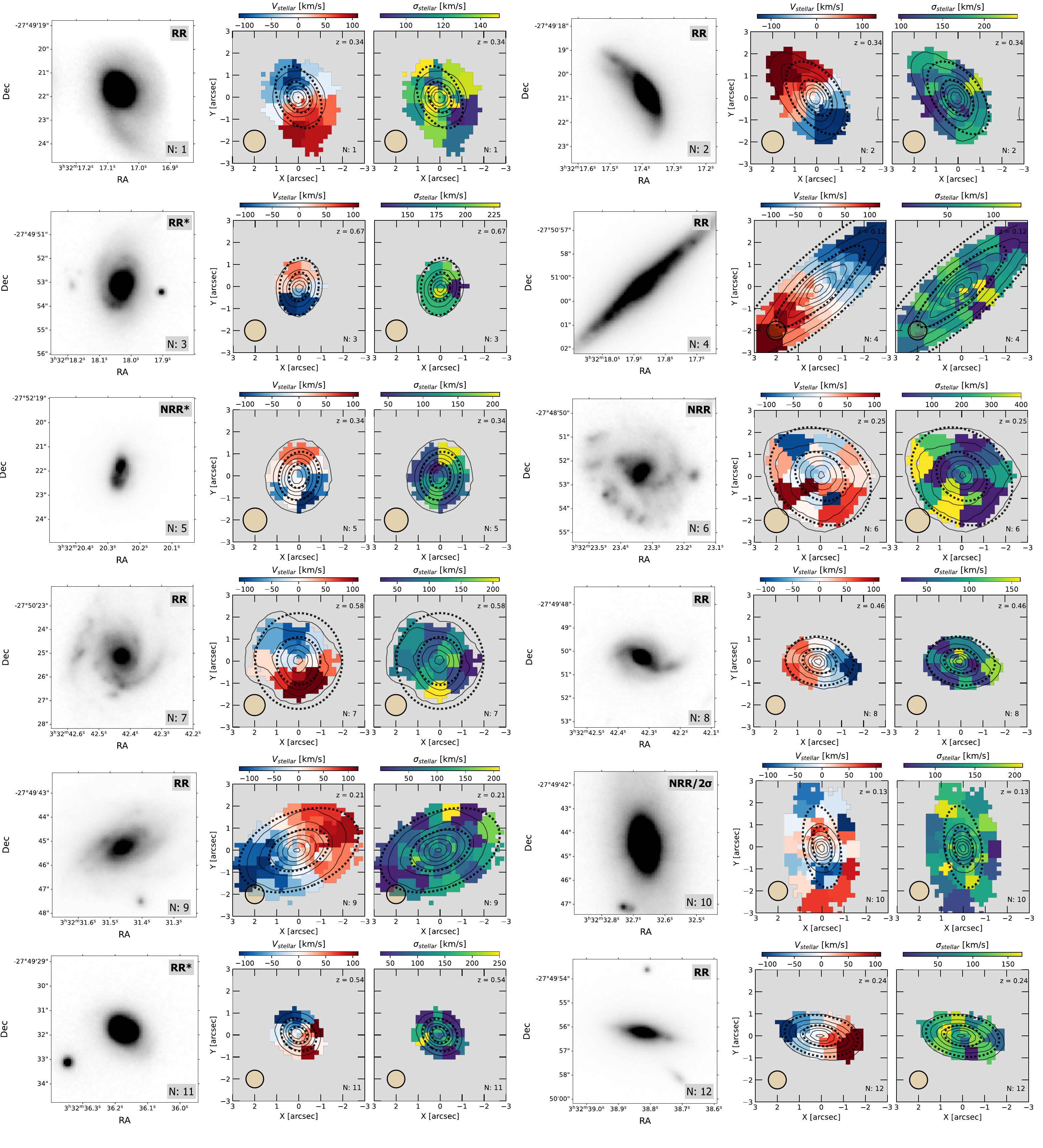}
        \caption{Resolved stellar velocity and velocity dispersion maps of the galaxy sample. Maps are Voronoi binned. The contours are isophotes of the surface brightness from the MUSE white light image plotted with black continuous lines. The dashed thick black lines show ellipses computed using 1$R_{e}$ and 2$R_{e}$, respectively. The number N in the lower right part of each panel denotes the galaxy index number found in  Table \ref{tab:sample_table}. Velocity and velocity dispersion ranges are colour-coded (see colour bar at the top of each   panel). The orange circles illustrate the size of the PSF (FWHM) in arcsec. The panels in greyscale correspond to the HST F160W image of the galaxy sample. The kinematic classification according to Table \ref{tab:sample_table} is included in the top right part of these panels. North is to the top and east is to the left.}
         \label{fig:maps1}
    \end{figure*}

\begin{figure*}
    \centering
       \includegraphics[width=19cm]{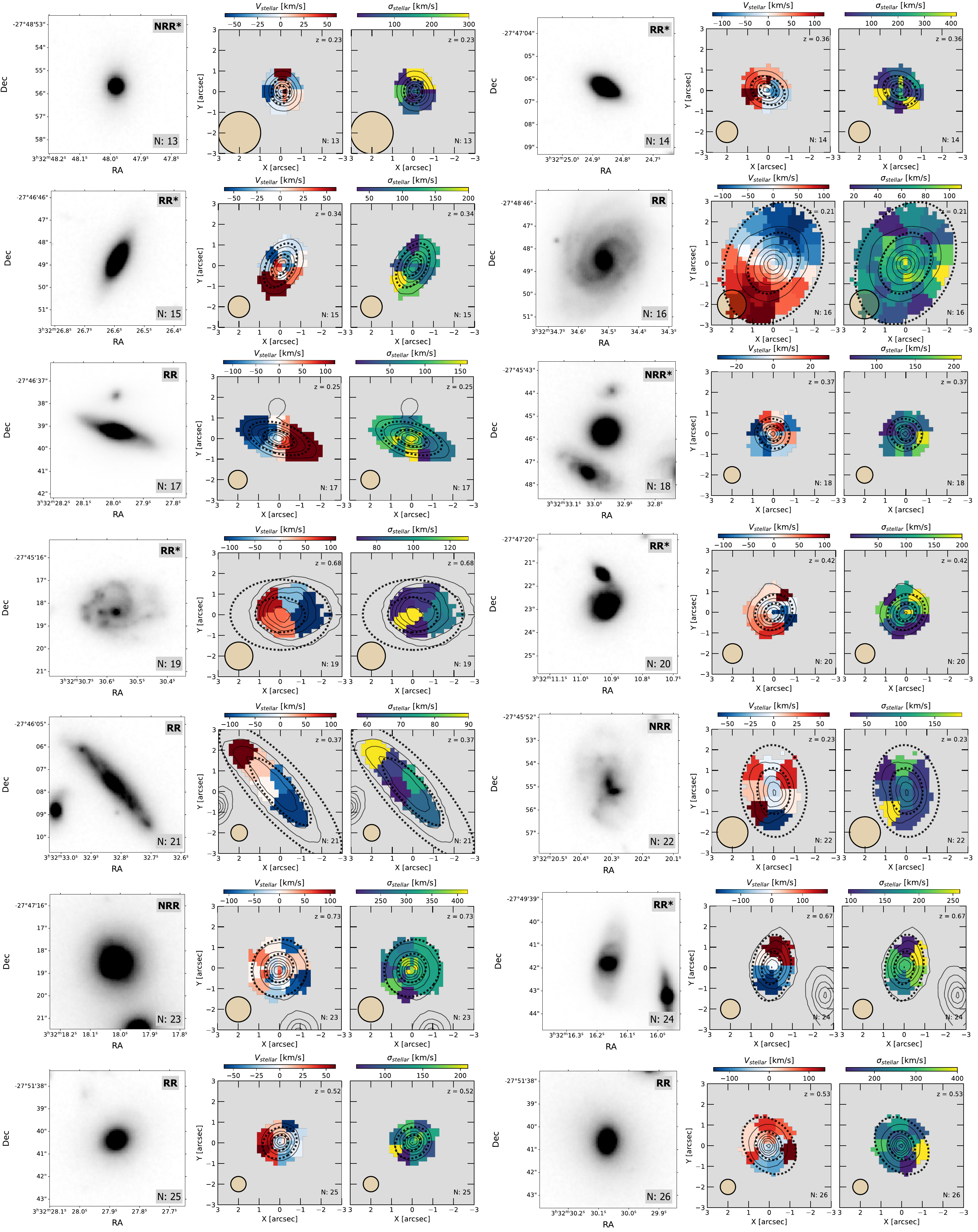}
         \caption{Same  as Fig. \ref{fig:maps1}.}
         \label{fig:maps2}
    \end{figure*}

\begin{figure*}
    \centering
       \includegraphics[width=19cm]{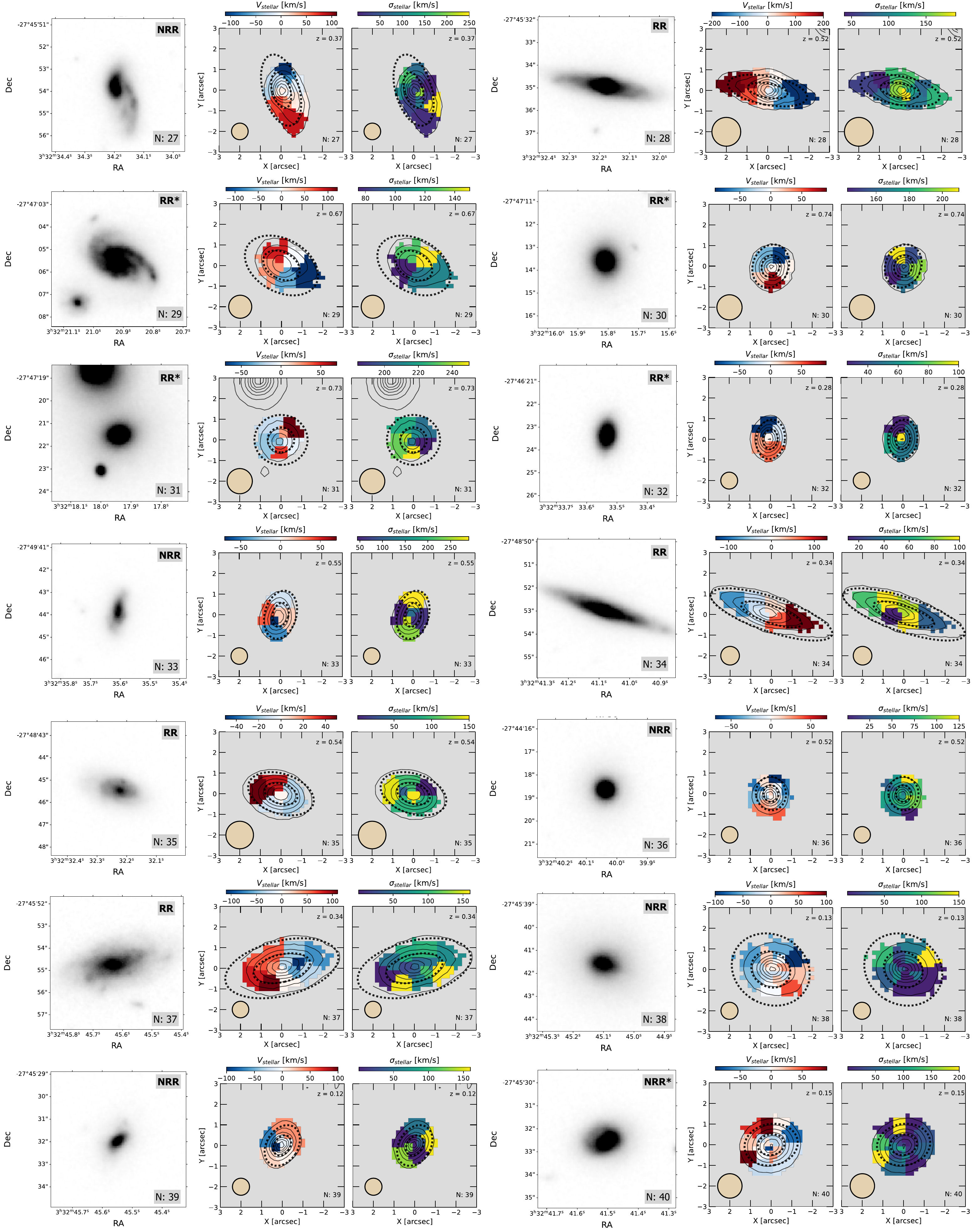}
         \caption{Same  as Fig. \ref{fig:maps1}.}
         \label{fig:maps3}
    \end{figure*}

\begin{figure*}
    \centering
       \includegraphics[width=19cm]{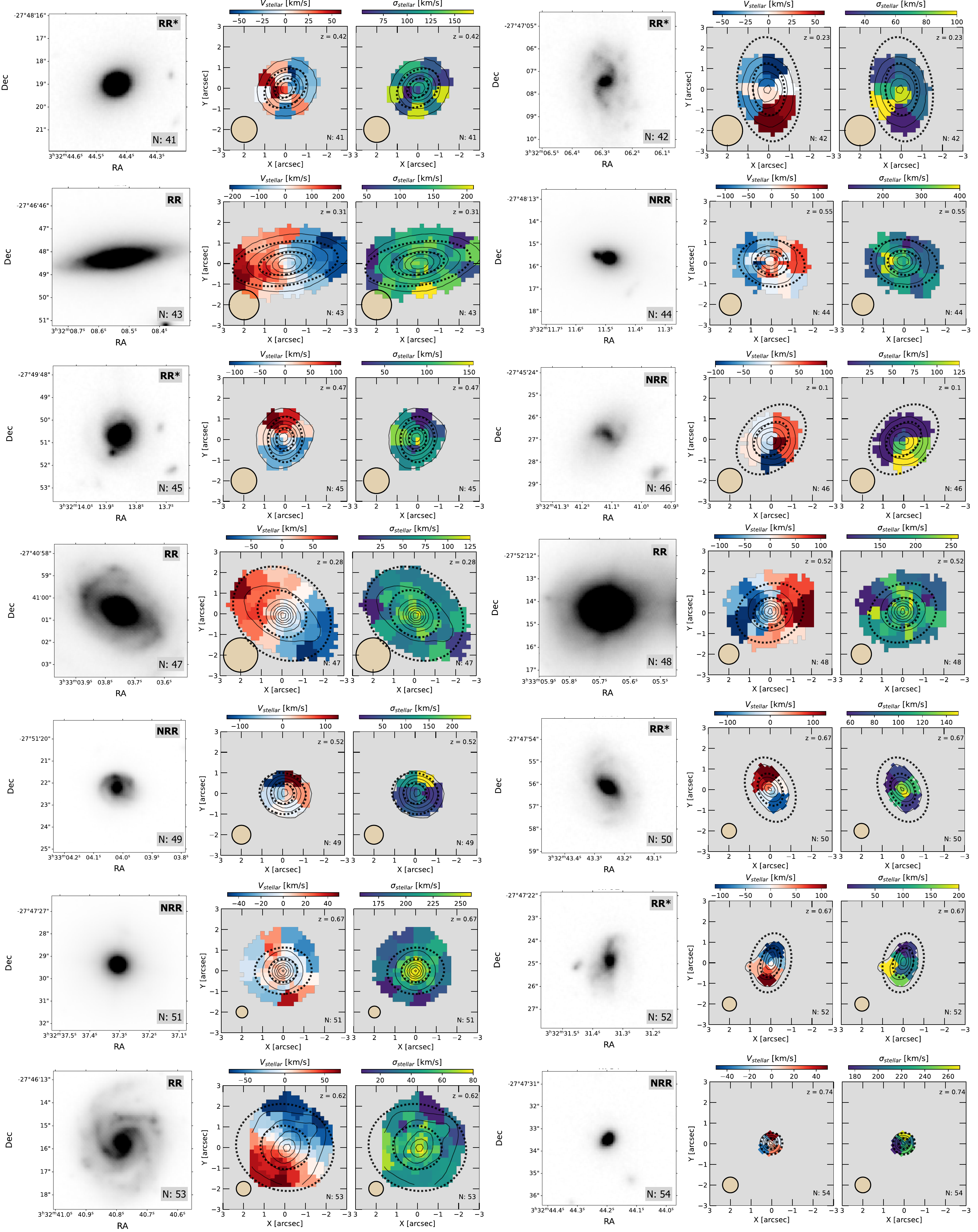}
         \caption{Same  as Fig. \ref{fig:maps1}.}
         \label{fig:maps4}
    \end{figure*}

\begin{figure*}
    \centering
       \includegraphics[width=19cm]{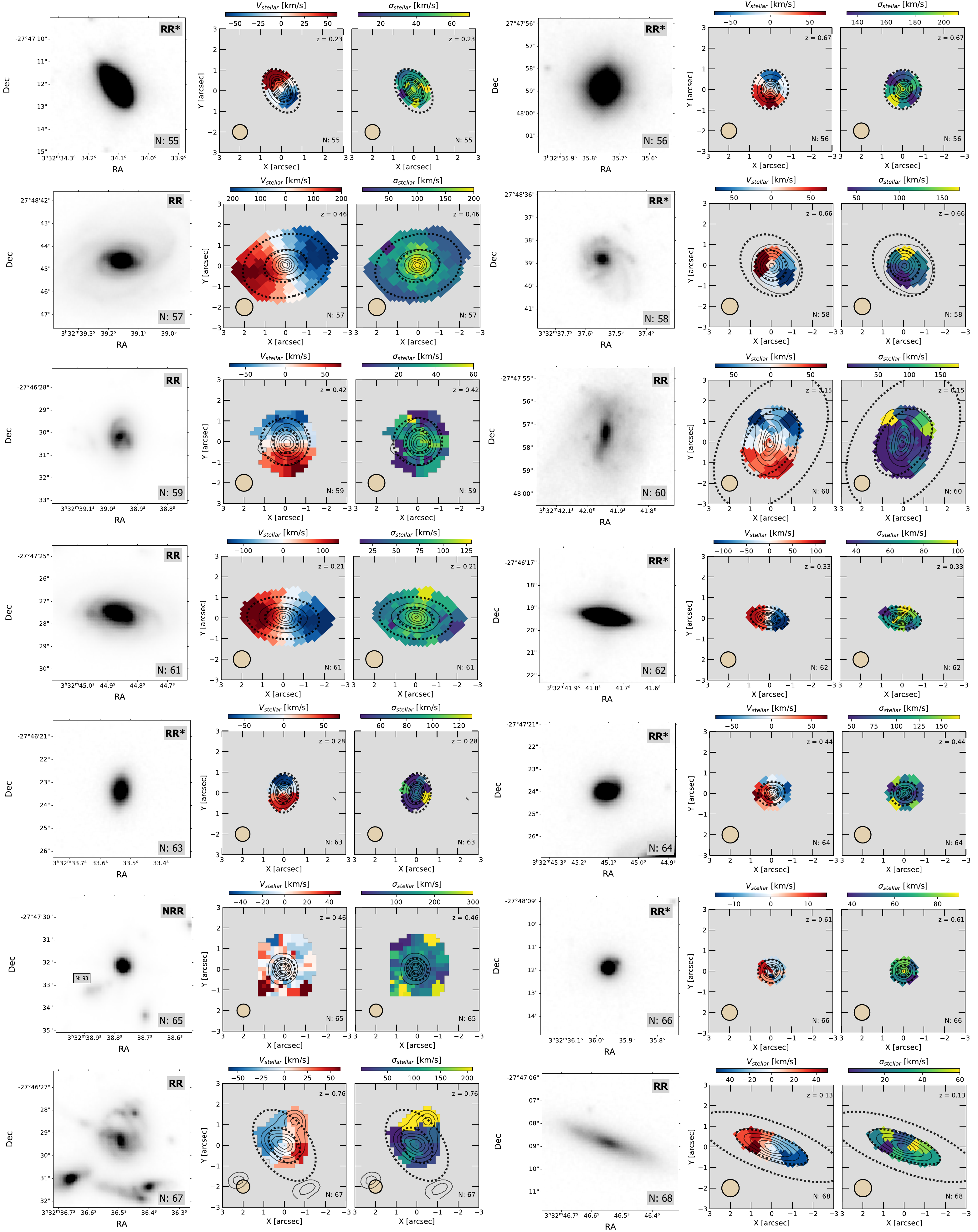}
         \caption{Same  as Fig. \ref{fig:maps1}.}
         \label{fig:maps5}
    \end{figure*}

\begin{figure*}
    \centering
       \includegraphics[width=19cm]{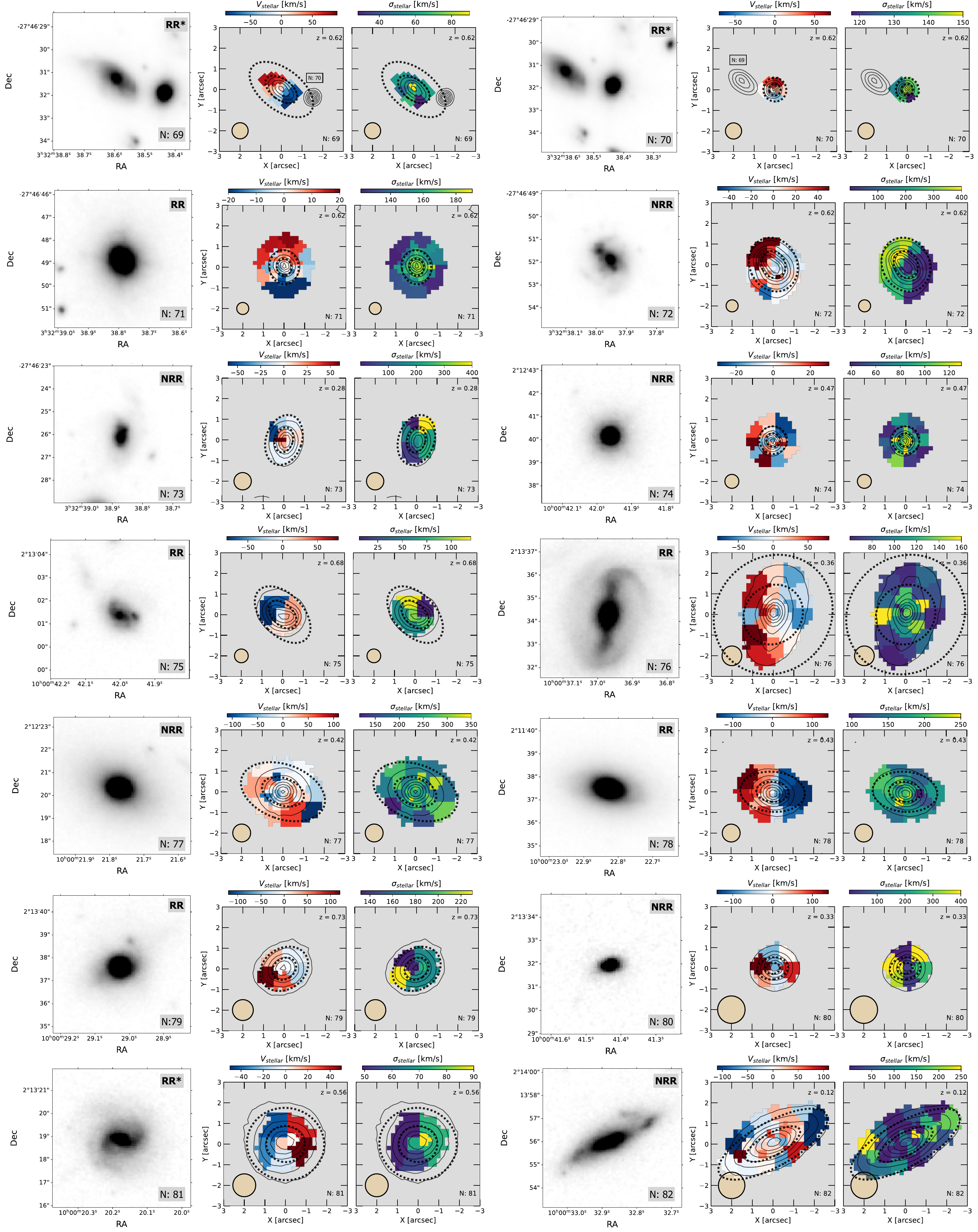}
         \caption{Same  as Fig. \ref{fig:maps1}.}
         \label{fig:maps6}
    \end{figure*}

\begin{figure*}
    \centering
       \includegraphics[width=19cm]{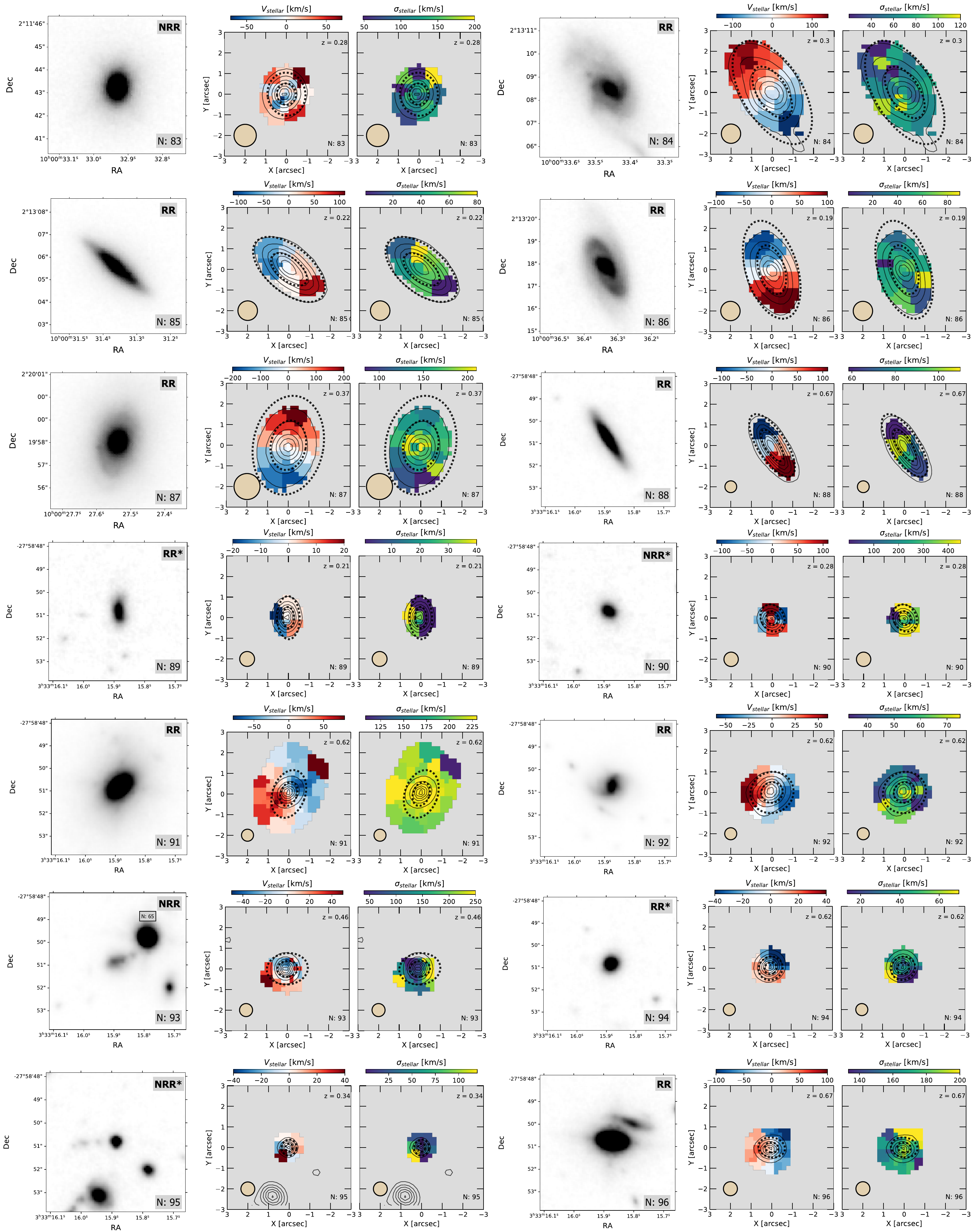}
         \caption{Same  as Fig. \ref{fig:maps1}.}
         \label{fig:maps7}
    \end{figure*}

\begin{figure*}
    \centering
       \includegraphics[width=19cm]{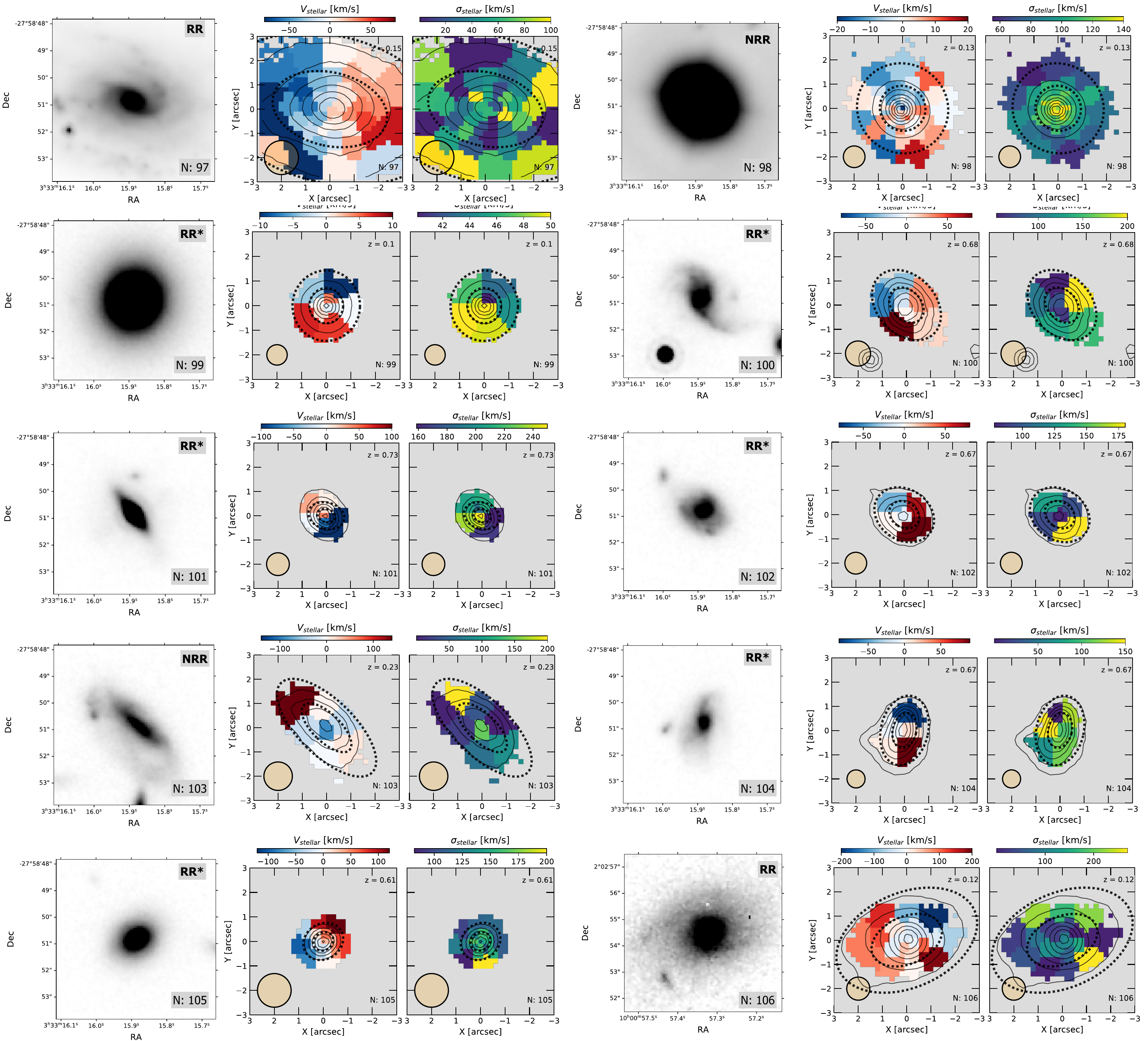}
        \caption{Same  as Fig. \ref{fig:maps1}.}
         \label{fig:maps8}
    \end{figure*}

Table \ref{tab:sample_table} contains the detailed photometric and spectroscopic information for the galaxy sample. 
From Fig. \ref{fig:maps1} to Fig. \ref{fig:maps8} photometric images and the kinematic maps (velocity and velocity dispersion) of the sample are presented, including a kinematic classification.

%%%%%%%%%%%%%%%%%%%%% Table %%%%%%%%%%%%%%%%%%%%%%%

\begin{landscape}
\small
\begin{longtable}{ccccccccccccccccc}   
\caption{Galaxy sample properties}\\
\hline 
    \hline
    \rule{0pt}{2ex} \hspace{0.1cm}N & ID & Field & z & RA & Dec & F160W & $R_{e}$ & $\sigma_{PSF}$ & $\epsilon$ & n & log$_{10}(M_{*})$ & log$_{10}$(SFR) & S/N & $\lambda_{R}$ & Aperture & KC \\ 
    \rule{0pt}{1ex}\hspace{0.15cm}- & - & - & - & ($^{\circ}$) & ($^{\circ}$) & (mag) & (arcsec) & (arcsec) & - & - & ($M_{\odot}$) & ($M_{\odot}$yr$^{-1}$) & - & ($R_{e}$) & - \\
    \vspace*{0.1cm}
    \rule{0pt}{1ex}\hspace{0.1cm}(1) & (2) & (3) & (4) & (5) & (6) & (7) & (8) & (9) & (10) & (11) & (12) & (13) & (14) & (15) & (16) & (17)  \\\hline 
    \endhead % all the lines above this will be repeated on every page
    1   & 17979 & GOODS-S & 0.338 & 53.07107  & -27.82274 & 18.5 & 0.59 & 0.42 & 0.4 & 3.8 & 10.55 & -0.54  & 27 & $\bullet$ 0.598 $\pm$ 0.004 & 2.0 & RR\\
    2   & 18117 & GOODS-S & 0.338 & 53.07249  & -27.82243 & 18.5 & 0.71 & 0.42 & 0.4 & 2.2 & 10.61 & 0.52   & 23 & $\bullet$ 0.823 $\pm$ 0.003 & 2.0 & RR\\
    3   & 16844 & GOODS-S & 0.670 & 53.07514  & -27.83142 & 19.3 & 0.54 & 0.41 & 0.3 & 2.0 & 10.81 & 0.32   & 14 & $\bullet$ 0.533 $\pm$ 0.006 & 2.0 & RR*\\
    4   & 11752 & GOODS-S & 0.122 & 53.07446  & -27.84980 & 17.4 & 1.4  & 0.36 & 0.7 & 1.2 & 9.96  & 0.76   & 37 & $\bullet$ 0.869 $\pm$ 0.003 & 2.0 & RR\\
    5   & 7336  & GOODS-S & 0.343 & 53.08442  & -27.87278 & 19.2 & 0.5  & 0.46 & 0.2 & 1.1 & 9.8   & 0.19   & 14 & $\bullet$ 0.728 $\pm$ 0.021 & 2.0 & NRR*\\
    6   & 20094 & GOODS-S & 0.247 & 53.09722  & -27.81458 & 19.8 & 1.12 & 0.47 & 0.2 & 2.8 & 9.2   & -1.97  & 13 & $\bullet$ 0.275 $\pm$ 0.010 & 2.0 & NRR\\
    7   & 14733 & GOODS-S & 0.577 & 53.17680  & -27.84034 & 19.4 & 1.04 & 0.41 & 0.1 & 2.9 & 10.66 & 1.53   & 16 & $\bullet$ 0.826 $\pm$ 0.008 & 1.5 & RR\\
    8   & 17138 & GOODS-S & 0.457 & 53.17634  & -27.83064 & 19.8 & 0.65 & 0.36 & 0.3 & 3.6 & 10.01 & -1.17  & 14 & $\bullet$ 0.830 $\pm$ 0.010 & 2.0 & RR\\
    9   & 17034 & GOODS-S & 0.214 & 53.13109  & -27.82924 & 18.4 & 1.1  & 0.36 & 0.4 & 1.7 & 10.19 & 0.24   & 33 & $\bullet$ 0.695 $\pm$ 0.006 & 2.0 & RR\\
    10  & 15773 & GOODS-S & 0.129 & 53.13610  & -27.82904 & 17.8 & 0.62 & 0.37 & 0.6 & 6.2 & 9.89  & -1.72  & 51 & $\bullet$ 0.327 $\pm$ 0.004 & 2.0 & NRR/2$\sigma$\\
    11  & 17928 & GOODS-S & 0.545 & 53.15073  & -27.82551 & 19.5 & 0.38 & 0.34 & 0.2 & 1.2 & 10.31 & 1.06   & 15 & $\bullet$ 0.528 $\pm$ 0.013 & 2.0 & RR*\\
    12  & 16635 & GOODS-S & 0.242 & 53.16176  & -27.83230 & 19.0 & 0.62 & 0.35 & 0.4 & 1.9 & 9.83  & 1.28   & 18 & $\bullet$ 0.660 $\pm$ 0.010 & 2.0 & RR\\
    13  & 19745 & GOODS-S & 0.233 & 53.19992  & -27.81547 & 20.2 & 0.26 & 0.87 & 0.3 & 6.4 & 9.29  & 0.05   & 15 & $\downarrow$ 0.316 $\pm$ 0.019 & 2.0 & NRR*\\
    14  & 27563 & GOODS-S & 0.363 & 53.10359  & -27.78510 & 19.9 & 0.37 & 0.44 & 0.3 & 0.8 & 9.56  & -0.74  & 16 & $\bullet$ 0.524 $\pm$ 0.016 & 2.0 & RR*\\
    15  & 28410 & GOODS-S & 0.337 & 53.11080  & -27.78021 & 19.5 & 0.44 & 0.44 & 0.4 & 1.3 & 10.12 & 0.53   & 14 & $\bullet$ 0.478 $\pm$ 0.010 & 2.0 & RR*\\
    16  & 19146 & GOODS-S & 0.215 & 53.14384  & -27.81347 & 18.0 & 1.33 & 0.59 & 0.3 & 2.6 & 10.38 & -0.08  & 35 & $\bullet$ 0.939 $\pm$ 0.003 & 2.0 & RR\\
    17  & 28888 & GOODS-S & 0.247 & 53.11662  & -27.77757 & 18.6 & 0.52 & 0.39 & 0.5 & 2.0 & 10.26 & 0.7    & 25 & $\bullet$ 0.939 $\pm$ 0.005 & 2.0 & RR\\
    18  & 32156 & GOODS-S & 0.365 & 53.13735  & -27.76269 & 19.3 & 0.37 & 0.34 & 0.1 & 5.4 & 10.38 & -1.4   & 21 & $\downarrow$ 0.623 $\pm$ 0.006 & 2.0 & NRR*\\
    19  & 33471 & GOODS-S & 0.681 & 53.12737  & -27.75507 & 19.9 & 1.01 & 0.58 & 0.3 & 0.7 & 10.26 & 1.02   & 14 & $\downarrow$ 0.708 $\pm$ 0.019 & 1.0 & RR*\\
    20  & 26491 & GOODS-S & 0.416 & 53.04549  & -27.78968 & 19.6 & 0.44 & 0.4  & 0.2 & 1.5 & 9.8   & 0.81   & 14 & $\bullet$ 0.779 $\pm$ 0.020 & 2.0 & RR*\\
    21  & 29723 & GOODS-S & 0.368 & 53.13674  & -27.76882 & 19.4 & 1.18 & 0.34 & 0.7 & 0.8 & 10.02 & 1.3    & 8  & $\downarrow$ 0.948 $\pm$ 0.004 & 1.0 & RR\\
    22  & 31246 & GOODS-S & 0.232 & 53.08464  & -27.76523 & 20.0 & 0.94 & 0.64 & 0.3 & 1.3 & 9.36  & -0.22  & 16 & $\bullet$ 0.462 $\pm$ 0.020 & 1.5 & NRR\\
    23  & 24878 & GOODS-S & 0.734 & 53.07506  & -27.78848 & 18.3 & 0.68 & 0.52 & 0.1 & 5.8 & 11.77 & -0.21  & 22 & $\bullet$ 0.447 $\pm$ 0.004 & 2.0 & NRR\\
    24  & 17223 & GOODS-S & 0.665 & 53.06733  & -27.82826 & 19.4 & 0.61 & 0.4  & 0.4 & 3.3 & 10.99 & 0.27   & 12 & $\bullet$ 0.818 $\pm$ 0.004 & 2.0 & RR*\\
    25  & 10176 & GOODS-S & 0.522 & 53.11616  & -27.86122 & 19.5 & 0.39 & 0.32 & 0.2 & 5.4 & 10.68 & -1.1   & 13 & $\downarrow$ 0.517 $\pm$ 0.006 & 2.0 & RR\\
    26  & 8813  & GOODS-S & 0.532 & 53.12530  & -27.86129 & 19.1 & 0.66 & 0.32 & 0.2 & 6.4 & 10.86 & -0.92  & 18 & $\bullet$ 0.369 $\pm$ 0.006 & 2.0 & RR\\
    27  & 31463 & GOODS-S & 0.366 & 53.14242  & -27.76504 & 20.3 & 0.62 & 0.34 & 0.5 & 1.3 & 9.3   & -1.01  & 13 & $\bullet$ 0.555 $\pm$ 0.023 & 1.5 & NRR\\
    28  & 32648 & GOODS-S & 0.524 & 53.13409  & -27.75970 & 19.6 & 0.58 & 0.59 & 0.6 & 3.8 & 10.68 & -0.03  & 14 & $\downarrow$ 0.844 $\pm$ 0.004 & 2.0 & RR\\
    29  & 27247 & GOODS-S & 0.668 & 53.08715  & -27.78485 & 19.4 & 0.78 & 0.47 & 0.3 & 0.9 & 10.62 & 0.82   & 11 & $\bullet$ 0.810 $\pm$ 0.011 & 1.5 & RR*\\
    30  & 26402 & GOODS-S & 0.738 & 53.06587  & -27.78711 & 19.7 & 0.48 & 0.52 & 0.2 & 3.7 & 10.9  & -0.71  & 12 & $\bullet$ 0.988 $\pm$ 0.005 & 2.0 & RR*\\
    31  & 26138 & GOODS-S & 0.733 & 53.07476  & -27.78930 & 19.3 & 0.62 & 0.52 & 0.1 & 7.1 & 11.23 & -0.75  & 13 & $\bullet$ 0.635 $\pm$ 0.005 & 2.0 & RR\\
    32  & 30082 & GOODS-S & 0.276 & 53.13973  & -27.77316 & 20.6 & 0.41 & 0.35 & 0.3 & 0.9 & 9.36  & -0.22  & 10 & $\bullet$ 0.841 $\pm$ 0.015 & 2.0 & RR*\\
    33  & 17361 & GOODS-S & 0.545 & 53.14838  & -27.82886 & 20.7 & 0.48 & 0.34 & 0.4 & 1.2 & 9.6   & 1.14   & 9  & $\bullet$ 0.199 $\pm$ 0.019 & 2.0 & NRR\\
    34  & 20356 & GOODS-S & 0.338 & 53.17118  & -27.81471 & 19.6 & 0.8  & 0.38 & 0.7 & 2.0 & 9.87  & 0.44   & 9  & $\downarrow$ 0.922 $\pm$ 0.004 & 1.5 & RR\\
    35  & 20898 & GOODS-S & 0.535 & 53.13430  & -27.81262 & 20.6 & 0.63 & 0.57 & 0.4 & 1.0 & 9.7   & 0.71   & 9  & $\bullet$ 0.597 $\pm$ 0.020 & 2.0 & RR\\
    36  & 36482 & GOODS-S & 0.522 & 53.16678  & -27.73854 & 20.1 & 0.39 & 0.34 & 0.1 & 2.8 & 10.14 & -0.21  & 13 & $\downarrow$ 0.685 $\pm$ 0.010 & 2.0 & NRR\\
    37  & 30918 & GOODS-S & 0.336 & 53.18986  & -27.76513 & 20.0 & 0.97 & 0.34 & 0.5 & 1.5 & 9.52  & -0.27  & 12 & $\bullet$ 0.844 $\pm$ 0.008 & 1.5 & RR\\
    38  & 31801 & GOODS-S & 0.127 & 53.18786  & -27.76151 & 19.8 & 0.91 & 0.35 & 0.1 & 1.8 & 8.5   & -1.59  & 19 & $\bullet$ 0.755 $\pm$ 0.013 & 1.5 & NRR\\
    39  & 32778 & GOODS-S & 0.124 & 53.18974  & -27.75886 & 21.1 & 0.45 & 0.35 & 0.3 & 1.9 & 8.09  & -3.08  & 11 & $\bullet$ 0.827 $\pm$ 0.031 & 2.0 & NRR\\
    40  & 32762 & GOODS-S & 0.147 & 53.17296  & -27.75900 & 20.0 & 0.57 & 0.5  & 0.2 & 0.9 & 8.97  & -0.42  & 15 & $\bullet$ 0.933 $\pm$ 0.015 & 2.0 & NRR*\\
    41  & 21978 & GOODS-S & 0.416 & 53.18512  & -27.80528 & 19.6 & 0.48 & 0.53 & 0.3 & 1.9 & 10.25 & 0.9    & 14 & $\bullet$ 0.494 $\pm$ 0.013 & 2.0 & RR*\\
    42  & 25622 & GOODS-S & 0.228 & 53.02618  & -27.78538 & 19.6 & 0.98 & 0.62 & 0.4 & 2.0 & 9.56  & -0.04  & 15 & $\bullet$ 0.765 $\pm$ 0.014 & 1.5 & RR*\\
    43  & 28098 & GOODS-S & 0.310 & 53.03544  & -27.78007 & 17.7 & 0.82 & 0.62 & 0.6 & 3.2 & 10.94 & 0.27   & 22 & $\downarrow$ 0.830 $\pm$ 0.002 & 2.0 & RR\\
    44  & 21832 & GOODS-S & 0.545 & 53.04781  & -27.80440 & 18.6 & 0.68 & 0.45 & 0.3 & 5.7 & 11.05 & -0.73  & 21 & $\downarrow$ 0.655 $\pm$ 0.003 & 2.0 & NRR\\
    45  & 16742 & GOODS-S & 0.468 & 53.05770  & -27.83078 & 19.6 & 0.48 & 0.53 & 0.3 & 2.5 & 10.47 & 0.36   & 11 & $\downarrow$ 0.819 $\pm$ 0.005 & 2.0 & RR*\\
    46  & 32855 & GOODS-S & 0.103 & 53.17136  & -27.75744 & 20.4 & 0.82 & 0.51 & 0.3 & 1.2 & 8.24  & -2.93  & 12 & $\bullet$ 0.678 $\pm$ 0.015 & 1.5 & NRR\\
    47  & 46386 & GOODS-S & 0.277 & 53.26568  & -27.68354 & 18.3 & 1.15 & 0.71 & 0.2 & 4.2 & 10.44 & -0.66  & 35 & $\bullet$ 0.822 $\pm$ 0.004 & 2.0 & RR\\
    48  & 6203  & GOODS-S & 0.519 & 53.27370  & -27.87071 & 18.0 & 0.76 & 0.42 & 0.2 & 5.7 & 11.23 & 0.14   & 33 & $\bullet$ 0.875 $\pm$ 0.003 & 2.0 & RR\\
    49  & 11536 & GOODS-S & 0.522 & 53.26672  & -27.85614 & 20.3 & 0.53 & 0.39 & 0.1 & 0.8 & 9.71  & 0.47   & 13 & $\bullet$ 0.948 $\pm$ 0.025 & 2.0 & NRR\\
    50  & 23802 & GOODS-S & 0.667 & 53.17998  & -27.79904 & 20.1 & 0.68 & 0.31 & 0.3 & 2.5 & 10.42 & -0.75  & 24 & $\bullet$ 0.997 $\pm$ 0.003 & 1.5 & RR*\\
    51  & 24622 & GOODS-S & 0.669 & 53.15545  & -27.79149 & 19.0 & 0.62 & 0.35 & 0.2 & 2.9 & 11.4  & -0.59  & 56 & $\bullet$ 0.322 $\pm$ 0.013 & 2.0 & NRR\\
    52  & 26322 & GOODS-S & 0.666 & 53.13065  & -27.79029 & 20.7 & 0.6  & 0.29 & 0.3 & 1.2 & 9.79  & 0.18   & 24 & $\bullet$ 0.845 $\pm$ 0.020 & 1.5 & RR*\\
    53  & 29845 & GOODS-S & 0.622 & 53.16954  & -27.77076 & 19.7 & 1.12 & 0.3  & 0.1 & 2.4 & 10.58 & 1.68   & 47 & $\bullet$ 0.858 $\pm$ 0.005 & 2.0 & RR\\
    54  & 25592 & GOODS-S & 0.737 & 53.18413  & -27.79264 & 21.2 & 0.26 & 0.32 & 0.0 & 2.1 & 9.77  & 0.95   & 19 & $\bullet$ 0.868 $\pm$ 0.040 & 2.0 & NRR\\
    55  & 26639 & GOODS-S & 0.227 & 53.14210  & -27.78668 & 20.2 & 0.48 & 0.31 & 0.3 & 0.6 & 9.02  & -0.78  & 42 & $\bullet$ 0.722 $\pm$ 0.023 & 2.0 & RR*\\
    56  & 23102 & GOODS-S & 0.665 & 53.14896  & -27.79969 & 19.5 & 0.46 & 0.32 & 0.1 & 4.5 & 10.99 & -0.79  & 36 & $\bullet$ 0.445 $\pm$ 0.003 & 2.0 & RR*\\
    57  & 20254 & GOODS-S & 0.457 & 53.16312  & -27.81238 & 18.4 & 0.89 & 0.35 & 0.3 & 7.2 & 10.94 & -0.16  & 62 & $\downarrow$ 0.732 $\pm$ 0.001 & 2.0 & RR\\
    58  & 20671 & GOODS-S & 0.665 & 53.15654  & -27.81066 & 20.6 & 0.74 & 0.34 & 0.2 & 1.5 & 9.83  & 0.59   & 23 & $\bullet$ 0.646 $\pm$ 0.023 & 1.0 & RR*\\
    59  & 29048 & GOODS-S & 0.419 & 53.16227  & -27.77504 & 20.3 & 0.58 & 0.35 & 0.0 & 1.1 & 9.46  & -0.15  & 38 & $\bullet$ 0.950 $\pm$ 0.010 & 2.0 & RR\\
    60  & 21890 & GOODS-S & 0.152 & 53.17476  & -27.79924 & 20.0 & 1.43 & 0.33 & 0.4 & 2.1 & 8.68  & -2.49  & 39 & $\downarrow$ 0.837 $\pm$ 0.012 & 1.0 & RR\\
    61  & 25172 & GOODS-S & 0.215 & 53.18722  & -27.79100 & 17.9 & 0.67 & 0.35 & 0.4 & 1.7 & 10.64 & 0.34   & 80 & $\downarrow$ 0.929 $\pm$ 0.003 & 2.0 & RR\\
    62  & 30133 & GOODS-S & 0.334 & 53.17396  & -27.77206 & 20.3 & 0.35 & 0.32 & 0.4 & 1.3 & 9.75  & 0.0    & 31 & $\downarrow$ 0.744 $\pm$ 0.006 & 2.0 & RR*\\
    63  & 30082 & GOODS-S & 0.276 & 53.13970  & -27.77316 & 20.6 & 0.41 & 0.3  & 0.3 & 0.9 & 9.36  & -0.22  & 26 & $\bullet$ 0.928 $\pm$ 0.009 & 2.0 & RR*\\
    64  & 26377 & GOODS-S & 0.435 & 53.18794  & -27.79003 & 19.8 & 0.31 & 0.34 & 0.1 & 2.1 & 9.84  & 0.23   & 38 & $\downarrow$ 0.747 $\pm$ 0.006 & 2.0 & RR*\\
    65  & 25909 & GOODS-S & 0.458 & 53.16156  & -27.79226 & 20.5 & 0.26 & 0.37 & 0.1 & 7.0 & 9.11  & 1.45   & 23 & $\bullet$ 0.681 $\pm$ 0.024 & 2.0 & NRR\\
    66  & 22942 & GOODS-S & 0.605 & 53.14984  & -27.80333 & 20.7 & 0.3  & 0.32 & 0.1 & 3.5 & 10.13 & -0.41  & 29 & $\downarrow$ 0.486 $\pm$ 0.006 & 2.0 & RR*\\
    67  & 28995 & GOODS-S & 0.765 & 53.15222  & -27.77476 & 21.0 & 0.77 & 0.27 & 0.4 & 3.1 & 9.83  & 0.59   & 25 & $\bullet$ 0.684 $\pm$ 0.027 & 1.5 & RR\\
    68  & 27438 & GOODS-S & 0.128 & 53.19407  & -27.78577 & 20.4 & 1.04 & 0.36 & 0.7 & 1.6 & 8.75  & -0.84  & 27 & $\downarrow$ 0.947 $\pm$ 0.003 & 1.0 & RR\\
    69  & 29238 & GOODS-S & 0.622 & 53.16069  & -27.77541 & 21.0 & 0.67 & 0.33 & 0.4 & 2.3 & 9.74  & -0.56  & 32 & $\bullet$ 0.955 $\pm$ 0.014 & 1.0 & RR*\\
    70  & 29794 & GOODS-S & 0.622 & 53.16016  & -27.77552 & 21.0 & 0.28 & 0.33 & 0.0 & 4.7 & 10.34 & -0.94  & 34 & $\downarrow$ 0.775 $\pm$ 0.004 & 2.0 & RR*\\
    71  & 27965 & GOODS-S & 0.621 & 53.16164  & -27.78025 & 20.4 & 0.38 & 0.36 & 0.1 & 7.0 & 10.53 & -1.25  & 57 & $\bullet$ 0.855 $\pm$ 0.017 & 2.0 & RR\\
    72  & 28445 & GOODS-S & 0.620 & 53.15818  & -27.78109 & 21.7 & 0.62 & 0.28 & 0.1 & 0.6 & 9.19  & 0.2    & 16 & $\bullet$ 0.183 $\pm$ 0.017 & 2.0 & NRR\\
    73  & 29861 & GOODS-S & 0.276 & 53.16190  & -27.77386 & 22.3 & 0.52 & 0.36 & 0.3 & 1.4 & 8.39  & -0.66  & 14 & $\bullet$ 0.132 $\pm$ 0.024 & 2.0 & NRR\\
    74  & 2957  & COSMOS  & 0.475 & 150.17485 & 2.21116   & 19.3 & 0.33 & 0.29 & 0.0 & 5.3 & 10.58 & 0.3    & 28 & $\bullet$ 0.493 $\pm$ 0.008 & 2.0 & NRR\\
    75  & 3487  & COSMOS  & 0.679 & 150.17502 & 2.21710   & 20.1 & 0.64 & 0.28 & 0.4 & 0.8 & 10.2  & 0.4    & 14 & $\bullet$ 0.918 $\pm$ 0.019 & 1.5 & RR\\
    76  & 4148  & COSMOS  & 0.361 & 150.15404 & 2.22625   & 18.3 & 1.44 & 0.4  & 0.1 & 3.9 & 10.69 & 0.41   & 17 & $\downarrow$ 0.680 $\pm$ 0.002 & 1.0 & RR\\
    77  & 1996  & COSMOS  & 0.424 & 150.09065 & 2.20565   & 18.2 & 0.83 & 0.37 & 0.4 & 7.5 & 11.12 & -0.86  & 25 & $\bullet$ 0.283 $\pm$ 0.005 & 2.0 & NRR\\
    78  & 902   & COSMOS  & 0.426 & 150.09505 & 2.19375   & 18.6 & 0.62 & 0.37 & 0.4 & 3.2 & 10.89 & 0.17   & 23 & $\bullet$ 0.879 $\pm$ 0.003 & 2.0 & RR\\
    79  & 4596  & COSMOS  & 0.732 & 150.12091 & 2.22718   &   -  & 0.55 & 0.43 & 0.2 & 3.0 & 11.09 & 0.19   & 12 & $\downarrow$ 0.631 $\pm$ 0.005 & 2.0 & RR\\
    80  & 4638  & COSMOS  & 0.325 & 150.17265 & 2.22551   & 21.1 & 0.32 & 0.56 & 0.3 & 1.3 & 8.75  & -0.49  & 8  & $\bullet$ 0.877 $\pm$ 0.023 & 2.0 & NRR\\
    81  & 4077  & COSMOS  & 0.561 & 150.08432 & 2.22182   & 19.6 & 0.91 & 0.47 & 0.1 & 1.6 & 10.77 & 0.12   & 11 & $\bullet$ 0.723 $\pm$ 0.009 & 1.5 & RR*\\
    82  & 5116  & COSMOS  & 0.123 & 150.13693 & 2.23230   & 19.6 & 0.89 & 0.54 & 0.6 & 1.1 & 8.6   & -1.19  & 18 & $\bullet$ 0.553 $\pm$ 0.025 & 2.0 & NRR\\
    83  & 901   & COSMOS  & 0.282 & 150.13724 & 2.19536   &   -  & 0.49 & 0.46 & 0.1 & 4.2 & 10.3  & 0.74   & 22 & $\bullet$ 0.574 $\pm$ 0.012 & 2.0 & NRR\\
    84  & 3165  & COSMOS  & 0.302 & 150.13943 & 2.21902   & 18.8 & 1.01 & 0.4  & 0.5 & 1.5 & 10.3  & 0.37   & 21 & $\bullet$ 0.802 $\pm$ 0.011 & 2.0 & RR\\
    85  & 3637  & COSMOS  & 0.219 & 150.13066 & 2.21821   & 19.2 & 0.76 & 0.4  & 0.5 & 0.8 & 9.79  & 0.37   & 17 & $\bullet$ 0.766 $\pm$ 0.010 & 2.0 & RR\\
    86  & 3880  & COSMOS  & 0.194 & 150.1514  & 2.22162   & 19.0 & 0.88 & 0.4  & 0.5 & 1.0 & 9.95  & 0.02   & 24 & $\bullet$ 0.837 $\pm$ 0.006 & 2.0 & RR\\
    87  & 14034 & COSMOS  & 0.373 & 150.11478 & 2.33266   & 17.9 & 1.01 & 0.54 & 0.3 & 3.8 & 10.93 & 0.39   & 30 & $\bullet$ 0.826 $\pm$ 0.003 & 2.0 & RR\\
    88  & 27564 & GOODS-S & 0.668 & 53.15723  & -27.78526 & 21.8 & 0.55 & 0.35 & 0.6 & 1.2 & 9.76  & -0.03  & 15 & $\bullet$ 0.820 $\pm$ 0.039 & 2.0 & RR\\
    89  & 29169 & GOODS-S & 0.214 & 53.16794  & -27.77808 & 23.4 & 0.38 & 0.32 & 0.5 & 4.9 & 7.95  & -1.51  & 20 & $\bullet$ 0.758 $\pm$ 0.039 & 2.0 & RR*\\
    90  & 27535 & GOODS-S & 0.276 & 53.17032  & -27.78526 & 23.6 & 0.31 & 0.32 & 0.2 & 2.3 & 7.97  & -1.17  & 12 & $\bullet$ 0.418 $\pm$ 0.038 & 2.0 & NRR*\\
    91  & 25053 & GOODS-S & 0.622 & 53.17252  & -27.78810 & 19.6 & 0.49 & 0.31 & 0.3 & 1.3 & 10.94 & -1.04  & 55 & $\bullet$ 0.279 $\pm$ 0.019 & 2.0 & RR\\
    92  & 28356 & GOODS-S & 0.620 & 53.17262  & -27.78097 & 20.9 & 0.51 & 0.3  & 0.1 & 2.9 & 9.77  & -0.53  & 42 & $\bullet$ 0.630 $\pm$ 0.025 & 2.0 & RR\\
    93  & 25487 & GOODS-S & 0.459 & 53.16196  & -27.79255 & 22.5 & 0.43 & 0.37 & 0.2 & 7.0 & 8.82  & -0.09  & 21 & $\bullet$ 0.362 $\pm$ 0.024 & 2.0 & NRR\\
    94  & 26233 & GOODS-S & 0.619 & 53.16802  & -27.78967 & 22.5 & 0.29 & 0.36 & 0.1 & 4.9 & 9.02  & 0.34   & 22 & $\downarrow$ 0.462 $\pm$ 0.041 & 2.0 & RR*\\
    95  & 25652 & GOODS-S & 0.337 & 53.16733  & -27.79186 & 23.4 & 0.19 & 0.37 & 0.1 & 4.7 & 8.3   & -1.1   & 20 & $\downarrow$ 0.299 $\pm$ 0.038 & 2.0 & NRR*\\
    96  & 25884 & GOODS-S & 0.666 & 53.16952  & -27.79197 & 20.2 & 0.28 & 0.38 & 0.3 & 1.6 & 10.6  & -0.68  & 17 & $\downarrow$ 0.584 $\pm$ 0.012 & 2.0 & RR\\
    97  & 9612  & GOODS-S & 0.148 & 53.25076  & -27.86122 & 18.1 & 1.91 & 0.6  & 0.4 & 4.8 & 7.48  & -2.98  & 36 & $\bullet$ 0.929 $\pm$ 0.015 & 1.5 & RR\\
    98  & 12252 & GOODS-S & 0.131 & 53.26336  & -27.84482 & 16.6 & 0.94 & 0.38 & 0.0 & 8.0 & 10.18 & -3.0   & 64 & $\downarrow$ 0.989 $\pm$ 0.001 & 2.0 & NRR\\
    99  & 36313 & GOODS-S & 0.104 & 53.18664  & -27.73507 & 17.7 & 0.7  & 0.35 & 0.0 & 7.0 & 9.68  & -0.41  & 46 & $\bullet$ 0.529 $\pm$ 0.024 & 2.0 & RR*\\
    100 & 10195 & GOODS-S & 0.679 & 53.08628  & -27.86173 & 19.9 & 0.7  & 0.44 & 0.3 & 0.9 & 10.08 & 1.44   & 10 & $\bullet$ 0.725 $\pm$ 0.022 & 2.0 & RR*\\
    101 & 20331 & GOODS-S & 0.735 & 53.08428  & -27.81404 & 19.4 & 0.32 & 0.4  & 0.3 & 3.1 & 10.87 &   -    & 13 & $\bullet$ 0.826 $\pm$ 0.033 & 2.0 & RR*\\
    102 & 9754  & GOODS-S & 0.672 & 53.10607  & -27.86519 & 20.1 & 0.58 & 0.38 & 0.2 & 1.0 & 10.15 & -0.15  & 9  & $\bullet$ 0.528 $\pm$ 0.025 & 2.0 & RR*\\
    103 & 17534 & GOODS-S & 0.232 & 53.11500  & -27.82725 & 20.2 & 0.87 & 0.5  & 0.6 & 1.2 & 9.26  & -0.12  & 13 & $\bullet$ 0.711 $\pm$ 0.012 & 1.5 & NRR\\
    104 & 26322 & GOODS-S & 0.666 & 53.13063  & -27.79025 & 20.6 & 0.6  & 0.32 & 0.3 & 3.2 & 9.79  & 0.18   & 10 & $\bullet$ 0.759 $\pm$ 0.030 & 2.0 & RR*\\
    105 & 35579 & GOODS-S & 0.606 & 53.13759  & -27.7435  & 19.6 & 0.39 & 0.58 & 0.1 & 1.6 & 10.63 &   -    & 19 & $\downarrow$ 0.675 $\pm$ 0.027 & 2.0 & RR*\\
    106 & 3181  & COSMOS  & 0.116 & 150.15631 & 2.21484   & 19.4 & 1.22 & 0.4  & 0.4 & 0.1 &  -    &   -    & 20 & $\bullet$ 0.666 $\pm$ 0.019 & 1.5 & RR\\ \hline
\caption{Column (1): galaxy index; column (2): 3D-HST galaxy unique identifier within a given field; column (3): field; column (4): MUSE spectroscopic redshift; column (5): RA J2000 degrees of galaxy centre from 3D-HST survey; column (6): Dec J2000 degrees of galaxy centre from 3D-HST survey; column (7): magnitude in the F160W filter from 3D-HST survey; column (8): galaxy effective radius; column (9): width of the Gaussian PSF (FWHM$_{\rm{PSF}}$/2.355); column (10): ellipticity; column (11): S\'{e}rsic index; column (12): galaxy stellar mass from 3D-HST survey; column (13): galaxy star formation rate from 3D-HST survey; column (14): MUSE signal-to-noise ratio of the galaxy measured within the effective radius; column (15): galaxy stellar angular momentum measured for the half-light ellipse aperture. Arrows ($\downarrow$) correspond to galaxies where the correction was not applied, and thus indicate measured  values, and filled dots ($\bullet$) indicate galaxies with corrected values; column (16): aperture in $R_{e}$ unit where $\lambda_{R}$ was computed; column (17): our kinematic classification (KC) according to the kinematics maps, RR stands for regular rotators, NRR for non-regular rotators, and RR* or NRR* are galaxies with clues of regular rotation or non-regular rotation, respectively, but are visually hard to classify.}
\label{tab:sample_table} 
\end{longtable}
\end{landscape}

%%%%%%%%%%%%%%%%%%%%%%%%%%%%%%%%%%%%%%%%%%%%%%%%%%%%%%%


%%%%%%%%%%%%%%%%%%%%%%%%%%%%%%%%%%%%%%%%%%%%%%%%%%%%%%%%%%%%%%%%%%%%%%%%%%%%%%%%%%%%%%%%%%%%%%%%%%%%%%
%\begin{figure*}[h]
% \vspace{-2cm}
% \wspace{-2cm}
% \includegraphics[scale=0.99]{aa59136e-26-1.pdf}
%\end{figure*}
%\begin{figure*}[h]
% \includegraphics[scale=0.99]{aa59136e-26-2.pdf}
%\end{figure*}
%\begin{figure*}[h]
% \includegraphics[scale=0.99]{aa59136e-26-3.pdf}
%\end{figure*}
%\begin{figure*}[h]
%    \includegraphics[scale=0.99]{aa59136e-26-4.pdf}
%\end{figure*}
\FloatBarrier
\includepdf[pages=-]{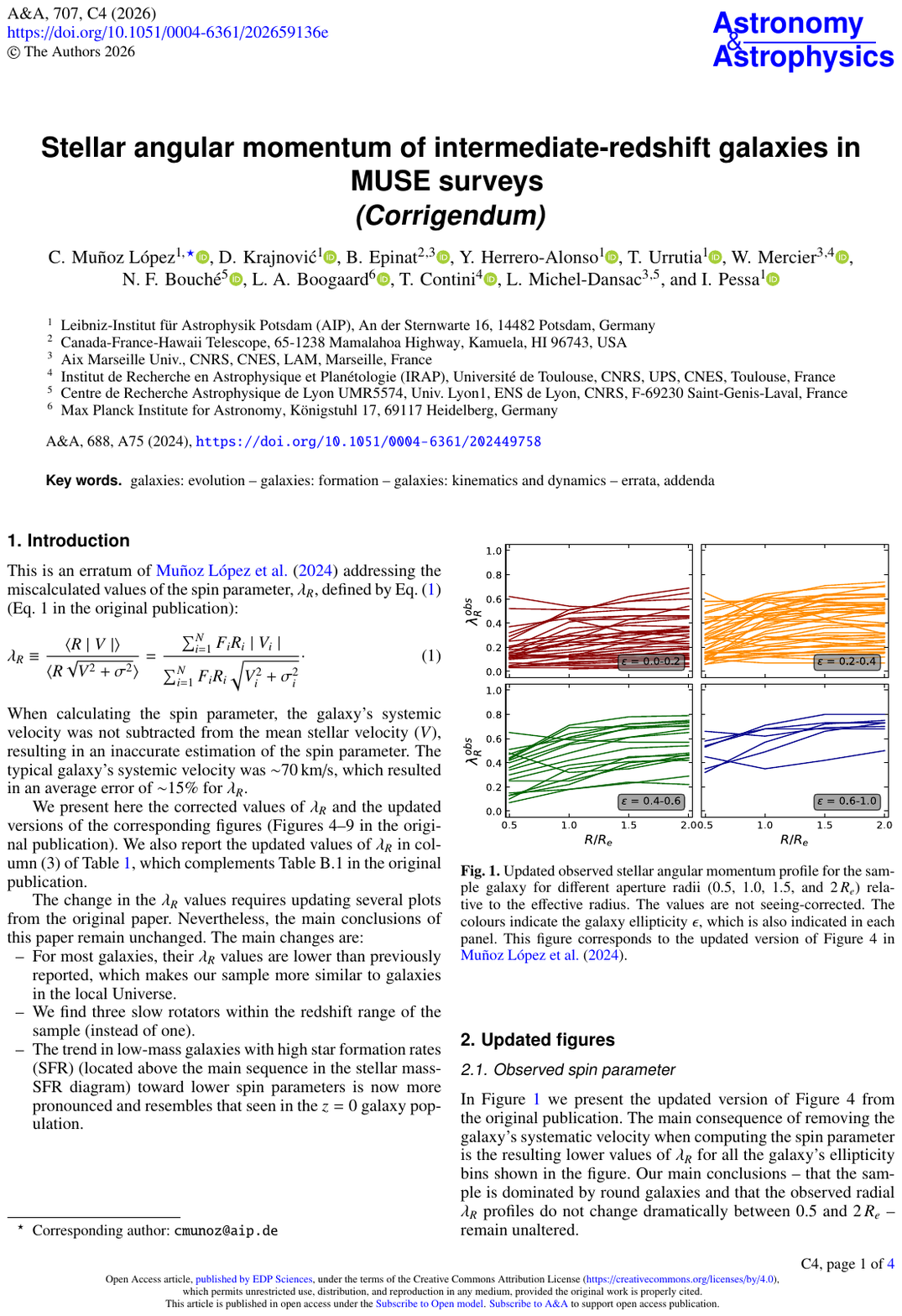}

\begin{thebibliography}{}
    \bibitem[Bacon et al.(2010)]{2010SPIE.7735E..08B} Bacon, R., Accardo, M., Adjali, L., et al.\ 2010, \procspie, 7735, 773508. %doi:10.1117/12.856027

    \bibitem[Bacon et al.(2017)]{2017A&A...608A...1B} Bacon, R., Conseil, S., Mary, D., et al.\ 2017, \aap, 608, A1. %doi:10.1051/0004-6361/201730833

    \bibitem[Bacon et al.(2023)]{2023A&A...670A...4B} Bacon, R., Brinchmann, J., Conseil, S., et al.\ 2023, \aap, 670, A4. %doi:10.1051/0004-6361/202244187

    \bibitem[Beckwith et al.(2006)]{2006AJ....132.1729B} Beckwith, S.~V.~W., Stiavelli, M., Koekemoer, A.~M., et al.\ 2006, \aj, 132, 1729. %doi:10.1086/507302

    \bibitem[Belli et al.(2017)]{2017ApJ...834...18B} Belli, S., Newman, A.~B., \& Ellis, R.~S.\ 2017, \apj, 834, 18. %doi:10.3847/1538-4357/834/1/18

    \bibitem[Beifiori et al.(2011)]{2011A&A...531A.109B} Beifiori, A., Maraston, C., Thomas, D., et al.\ 2011, \aap, 531, A109. %doi:10.1051/0004-6361/201016323

    \bibitem[Bezanson et al.(2018)]{2018ApJ...858...60B} Bezanson, R., van der Wel, A., Pacifici, C., et al.\ 2018, \apj, 858, 60. %doi:10.3847/1538-4357/aabc55

    \bibitem[Bouwens et al.(2011)]{2011ApJ...737...90B} Bouwens, R.~J., Illingworth, G.~D., Oesch, P.~A., et al.\ 2011, \apj, 737, 90. %doi:10.1088/0004-637X/737/2/90

    \bibitem[Brammer et al.(2012)]{2012ApJS..200...13B} Brammer, G.~B., van Dokkum, P.~G., Franx, M., et al.\ 2012, \apjs, 200, 13. %doi:10.1088/0067-0049/200/2/13

    \bibitem[Brough et al.(2017)]{2017ApJ...844...59B} Brough, S., van de Sande, J., Owers, M.~S., et al.\ 2017, \apj, 844, 59. %doi:10.3847/1538-4357/aa7a11

    \bibitem[Buck et al.(2019)]{2019ApJ...874...67B} Buck, T., Ness, M., Obreja, A., et al.\ 2019, \apj, 874, 67. %doi:10.3847/1538-4357/aaffd0

    \bibitem[Bundy et al.(2015)]{2015ApJ...798....7B} Bundy, K., Bershady, M.~A., Law, D.~R., et al.\ 2015, \apj, 798, 7. %doi:10.1088/0004-637X/798/1/7

    \bibitem[Cameron et al.(2009)]{2009ApJ...699..105C} Cameron, E., Driver, S.~P., Graham, A.~W., et al.\ 2009, \apj, 699, 105. %doi:10.1088/0004-637X/699/1/105

    \bibitem[Cappellari \& Copin(2003)]{2003MNRAS.342..345C} Cappellari, M. \& Copin, Y.\ 2003, \mnras, 342, 345. %doi:10.1046/j.1365-8711.2003.06541.x

    \bibitem[Cappellari \& Emsellem(2004)]{2004PASP..116..138C} Cappellari, M. \& Emsellem, E.\ 2004, \pasp, 116, 138. %doi:10.1086/381875

    \bibitem[Cappellari et al.(2007)]{2007MNRAS.379..418C} Cappellari, M., Emsellem, E., Bacon, R., et al.\ 2007, \mnras, 379, 418. %doi:10.1111/j.1365-2966.2007.11963.x

    \bibitem[Cappellari(2008)]{2008MNRAS.390...71C} Cappellari, M.\ 2008, \mnras, 390, 71. %doi:10.1111/j.1365-2966.2008.13754.x
    
    \bibitem[Cappellari et al.(2011)]{2011MNRAS.413..813C} Cappellari, M., Emsellem, E., Krajnovi{\'c}, D., et al.\ 2011, \mnras, 413, 813. %doi:10.1111/j.1365-2966.2010.18174.x

    \bibitem[Cappellari et al.(2011)]{2011MNRAS.416.1680C} Cappellari, M., Emsellem, E., Krajnovi{\'c}, D., et al.\ 2011, \mnras, 416, 1680. %doi:10.1111/j.1365-2966.2011.18600.x

    \bibitem[Cappellari et al.(2013)]{2013MNRAS.432.1862C} Cappellari, M., McDermid, R.~M., Alatalo, K., et al.\ 2013, \mnras, 432, 1862. %doi:10.1093/mnras/stt644
    
    \bibitem[Cappellari(2016)]{2016ARA&A..54..597C} Cappellari, M.\ 2016, \araa, 54, 597. %doi:10.1146/annurev-astro-082214-122432

    \bibitem[Cappellari(2017)]{2017MNRAS.466..798C} Cappellari, M.\ 2017, \mnras, 466, 798. %doi:10.1093/mnras/stw3020

    \bibitem[Croom et al.(2012)]{2012MNRAS.421..872C} Croom, S.~M., Lawrence, J.~S., Bland-Hawthorn, J., et al.\ 2012, \mnras, 421, 872. %doi:10.1111/j.1365-2966.2011.20365.x

    \bibitem[Croom et al.(2024)]{2024arXiv240206877C} Croom, S.~M., van de Sande, J., Vaughan, S.~P., et al.\ 2024.% arXiv:2402.06877. %doi:10.48550/arXiv.2402.06877

    \bibitem[Davis et al.(1985)]{1985ApJ...292..371D} Davis, M., Efstathiou, G., Frenk, C.~S., et al.\ 1985, \apj, 292, 371. %doi:10.1086/163168

    \bibitem[D'Eugenio et al.(2013)]{2013MNRAS.429.1258D} D'Eugenio, F., Houghton, R.~C.~W., Davies, R.~L., et al.\ 2013, \mnras, 429, 1258. %doi:10.1093/mnras/sts406

    \bibitem[De Lucia et al.(2012)]{2012MNRAS.423.1277D} De Lucia, G., Weinmann, S., Poggianti, B.~M., et al.\ 2012, \mnras, 423, 1277. %doi:10.1111/j.1365-2966.2012.20983.x

    \bibitem[Emsellem et al.(2007)]{2007MNRAS.379..401E} Emsellem, E., Cappellari, M., Krajnovi{\'c}, D., et al.\ 2007, \mnras, 379, 401. %doi:10.1111/j.1365-2966.2007.11752.x

    \bibitem[Emsellem et al.(2011)]{2011MNRAS.414..888E} Emsellem, E., Cappellari, M., Krajnovi{\'c}, D., et al.\ 2011, \mnras, 414, 888. %doi:10.1111/j.1365-2966.2011.18496.x
    
    \bibitem[Epinat et al.(2023)]{2023arXiv231200924E} Epinat, B., Contini, T., Mercier, W., et al.\ 2023.% arXiv:2312.00924. %doi:10.48550/arXiv.2312.00924

    \bibitem[Franx \& van Dokkum(2001)]{2001ASPC..230..581F} Franx, M. \& van Dokkum, P.~G.\ 2001, Galaxy Disks and Disk Galaxies, 230, 581

    \bibitem[F{\"o}rster Schreiber et al.(2011)]{2011ApJ...739...45F} F{\"o}rster Schreiber, N.~M., Shapley, A.~E., Genzel, R., et al.\ 2011, \apj, 739, 45. %doi:10.1088/0004-637X/739/1/45

    \bibitem[Fraser-McKelvie et al.(2021)]{2021MNRAS.503.4992F} Fraser-McKelvie, A., Cortese, L., van de Sande, J., et al.\ 2021, \mnras, 503, 4992. %doi:10.1093/mnras/stab573

    \bibitem[Gadotti et al.(2020)]{2020A&A...643A..14G} Gadotti, D.~A., Bittner, A., Falc{\'o}n-Barroso, J., et al.\ 2020, \aap, 643, A14. %doi:10.1051/0004-6361/202038448

    \bibitem[Gerhard(1993)]{1993MNRAS.265..213G} Gerhard, O.~E.\ 1993, \mnras, 265, 213. %doi:10.1093/mnras/265.1.213

    \bibitem[Graham et al.(2018)]{2018MNRAS.477.4711G} Graham, M.~T., Cappellari, M., Li, H., et al.\ 2018, \mnras, 477, 4711. %doi:10.1093/mnras/sty504

    \bibitem[Greene et al.(2017)]{2017ApJ...851L..33G} Greene, J.~E., Leauthaud, A., Emsellem, E., et al.\ 2017, \apjl, 851, L33. %doi:10.3847/2041-8213/aa8ace

    \bibitem[Grogin et al.(2011)]{2011ApJS..197...35G} Grogin, N.~A., Kocevski, D.~D., Faber, S.~M., et al.\ 2011, \apjs, 197, 35. %doi:10.1088/0067-0049/197/2/35

    \bibitem[Gu{\'e}rou et al.(2017)]{2017A&A...608A...5G} Gu{\'e}rou, A., Krajnovi{\'c}, D., Epinat, B., et al.\ 2017, \aap, 608, A5. %doi:10.1051/0004-6361/201730905

    \bibitem[Greene et al.(2017)]{2017ApJ...851L..33G} Greene, J.~E., Leauthaud, A., Emsellem, E., et al.\ 2017, \apjl, 851, L33. %doi:10.3847/2041-8213/aa8ace

    \bibitem[Greene et al.(2018)]{2018ApJ...852...36G} Greene, J.~E., Leauthaud, A., Emsellem, E., et al.\ 2018, \apj, 852, 36.% doi:10.3847/1538-4357/aa9bde

    \bibitem[Harborne et al.(2020)]{2020MNRAS.497.2018H} Harborne, K.~E., van de Sande, J., Cortese, L., et al.\ 2020, \mnras, 497, 2018. %doi:10.1093/mnras/staa1847

    \bibitem[Hopkins \& Beacom(2006)]{2006ApJ...651..142H} Hopkins, A.~M. \& Beacom, J.~F.\ 2006, \apj, 651, 142. %doi:10.1086/506610

    \bibitem[Hopkins et al.(2009)]{2009MNRAS.398..898H} Hopkins, P.~F., Bundy, K., Murray, N., et al.\ 2009, \mnras, 398, 898. %doi:10.1111/j.1365-2966.2009.15062.x

    \bibitem[Krajnovi{\'c} et al.(2006)]{2006MNRAS.366..787K} Krajnovi{\'c}, D., Cappellari, M., de Zeeuw, P.~T., et al.\ 2006, \mnras, 366, 787. %doi:10.1111/j.1365-2966.2005.09902.x

    \bibitem[Krajnovi{\'c} et al.(2008)]{2008MNRAS.390...93K} Krajnovi{\'c}, D., Bacon, R., Cappellari, M., et al.\ 2008, \mnras, 390, 93. %doi:10.1111/j.1365-2966.2008.13712.x

    \bibitem[Krajnovi{\'c} et al.(2011)]{2011MNRAS.414.2923K} Krajnovi{\'c}, D., Emsellem, E., Cappellari, M., et al.\ 2011, \mnras, 414, 2923. %doi:10.1111/j.1365-2966.2011.18560.x

    \bibitem[Krajnovi{\'c} et al.(2013)]{2013MNRAS.432.1768K} Krajnovi{\'c}, D., Alatalo, K., Blitz, L., et al.\ 2013, \mnras, 432, 1768. %doi:10.1093/mnras/sts315

    \bibitem[Krajnovi{\'c} et al.(2020)]{2020A&A...635A.129K} Krajnovi{\'c}, D., Ural, U., Kuntschner, H., et al.\ 2020, \aap, 635, A129. %doi:10.1051/0004-6361/201937040

    \bibitem[Kriek et al.(2009)]{2009ApJ...700..221K} Kriek, M., van Dokkum, P.~G., Labb{\'e}, I., et al.\ 2009, \apj, 700, 221. %doi:10.1088/0004-637X/700/1/221

    \bibitem[Khochfar et al.(2011)]{2011MNRAS.417..845K} Khochfar, S., Emsellem, E., Serra, P., et al.\ 2011, \mnras, 417, 845. %doi:10.1111/j.1365-2966.2011.19486.x

    \bibitem[Lagos et al.(2018)]{2018MNRAS.476.4327L} Lagos, C. del P., Schaye, J., Bah{\'e}, Y., et al.\ 2018, \mnras, 476, 4327. %doi:10.1093/mnras/sty489

    \bibitem[Lagos et al.(2022)]{2022MNRAS.509.4372L} Lagos, C. del P., Emsellem, E., van de Sande, J., et al.\ 2022, \mnras, 509, 4372. %doi:10.1093/mnras/stab3128

    \bibitem[Ma et al.(2014)]{2014ApJ...795..158M} Ma, C.-P., Greene, J.~E., McConnell, N., et al.\ 2014, \apj, 795, 158. %doi:10.1088/0004-637X/795/2/158
    
    \bibitem[Madau \& Dickinson(2014)]{2014ARA&A..52..415M} Madau, P. \& Dickinson, M.\ 2014, \araa, 52, 415. %doi:10.1146/annurev-astro-081811-125615

    \bibitem[Moffat(1969)]{1969A&A.....3..455M} Moffat, A.~F.~J.\ 1969, \aap, 3, 455

    \bibitem[Martig et al.(2009)]{2009ApJ...707..250M} Martig, M., Bournaud, F., Teyssier, R., et al.\ 2009, \apj, 707, 250. %doi:10.1088/0004-637X/707/1/250
    
    \bibitem[Naab et al.(2009)]{2009ApJ...699L.178N} Naab, T., Johansson, P.~H., \& Ostriker, J.~P.\ 2009, \apjl, 699, L178. %doi:10.1088/0004-637X/699/2/L178
    
    \bibitem[Noble et al.(2013)]{2013MNRAS.436L..40N} Noble, A.~G., Geach, J.~E., van Engelen, A.~J., et al.\ 2013, \mnras, 436, L40. %doi:10.1093/mnrasl/slt108

    \bibitem[Noble et al.(2013)]{2013ApJ...768..118N} Noble, A.~G., Webb, T.~M.~A., Muzzin, A., et al.\ 2013, \apj, 768, 118. %doi:10.1088/0004-637X/768/2/118
    
    \bibitem[Ramella et al.(2001)]{2001A&A...368..776R} Ramella, M., Boschin, W., Fadda, D., et al.\ 2001, \aap, 368, 776. %doi:10.1051/0004-6361:20010071

    \bibitem[S{\'a}nchez et al.(2012)]{2012A&A...538A...8S} S{\'a}nchez, S.~F., Kennicutt, R.~C., Gil de Paz, A., et al.\ 2012, \aap, 538, A8. %doi:10.1051/0004-6361/201117353

    \bibitem[Schulze et al.(2018)]{2018MNRAS.480.4636S} Schulze, F., Remus, R.-S., Dolag, K., et al.\ 2018, \mnras, 480, 4636. %doi:10.1093/mnras/sty2090

    \bibitem[Scoville et al.(2007)]{2007ApJS..172....1S} Scoville, N., Aussel, H., Brusa, M., et al.\ 2007, \apjs, 172, 1. %doi:10.1086/516585

    \bibitem[Skelton et al.(2014)]{2014ApJS..214...24S} Skelton, R.~E., Whitaker, K.~E., Momcheva, I.~G., et al.\ 2014, \apjs, 214, 24. %doi:10.1088/0067-0049/214/2/24

    \bibitem[Shi et al.(2021)]{2021ApJ...911...46S} Shi, K., Toshikawa, J., Lee, K.-S., et al.\ 2021, \apj, 911, 46. %doi:10.3847/1538-4357/abe62e
    
    \bibitem[Simons et al.(2017)]{2017ApJ...843...46S} Simons, R.~C., Kassin, S.~A., Weiner, B.~J., et al.\ 2017, \apj, 843, 46. %doi:10.3847/1538-4357/aa740c
    
    \bibitem[Soto et al.(2016)]{2016MNRAS.458.3210S} Soto, K.~T., Lilly, S.~J., Bacon, R., et al.\ 2016, \mnras, 458, 3210. %doi:10.1093/mnras/stw474

    \bibitem[Stott et al.(2016)]{2016MNRAS.457.1888S} Stott, J.~P., Swinbank, A.~M., Johnson, H.~L., et al.\ 2016, \mnras, 457, 1888. %doi:10.1093/mnras/stw129

    \bibitem[Tacconi et al.(2013)]{2013ApJ...768...74T} Tacconi, L.~J., Neri, R., Genzel, R., et al.\ 2013, \apj, 768, 74. %doi:10.1088/0004-637X/768/1/74

    \bibitem[Urrutia et al.(2019)]{2019A&A...624A.141U} Urrutia, T., Wisotzki, L., Kerutt, J., et al.\ 2019, \aap, 624, A141. %doi:10.1051/0004-6361/201834656

    \bibitem[Valdes et al.(2004)]{2004ApJS..152..251V} Valdes, F., Gupta, R., Rose, J.~A., et al.\ 2004, \apjs, 152, 251. %doi:10.1086/386343

    \bibitem[van der Marel \& Franx(1993)]{1993ApJ...407..525V} van der Marel, R.~P. \& Franx, M.\ 1993, \apj, 407, 525. %doi:10.1086/172534

    \bibitem[van der Marel \& van Dokkum(2007)]{2007ApJ...668..738V} van der Marel, R.~P. \& van Dokkum, P.~G.\ 2007, \apj, 668, 738. %doi:10.1086/521210

    \bibitem[van de Sande et al.(2013)]{2013ApJ...771...85V} van de Sande, J., Kriek, M., Franx, M., et al.\ 2013, \apj, 771, 85. %doi:10.1088/0004-637X/771/2/85

    \bibitem[van de Sande et al.(2021)]{2021MNRAS.505.3078V} van de Sande, J., Vaughan, S.~P., Cortese, L., et al.\ 2021, \mnras, 505, 3078. %doi:10.1093/mnras/stab1490

    \bibitem[van de Sande et al.(2021)]{2021MNRAS.508.2307V} van de Sande, J., Croom, S.~M., Bland-Hawthorn, J., et al.\ 2021, \mnras, 508, 2307. %doi:10.1093/mnras/stab2647

    \bibitem[van der Wel \& van der Marel(2008)]{2008ApJ...684..260V} van der Wel, A. \& van der Marel, R.~P.\ 2008, \apj, 684, 260. %doi:10.1086/589734

    \bibitem[van der Wel et al.(2014)]{2014ApJ...792L...6V} van der Wel, A., Chang, Y.-Y., Bell, E.~F., et al.\ 2014, \apjl, 792, L6. %doi:10.1088/2041-8205/792/1/L6

    \bibitem[van Dokkum et al.(2011)]{2011ApJ...743L..15V} van Dokkum, P.~G., Brammer, G., Fumagalli, M., et al.\ 2011, \apjl, 743, L15. %doi:10.1088/2041-8205/743/1/L15

    \bibitem[Vaughan et al.(2024)]{2024MNRAS.528.5852V} Vaughan, S.~P., van de Sande, J., Fraser-McKelvie, A., et al.\ 2024, \mnras, 528, 5852. %doi:10.1093/mnras/stae409

    \bibitem[Wang et al.(2020)]{2020MNRAS.495.1958W} Wang, B., Cappellari, M., Peng, Y., et al.\ 2020, \mnras, 495, 1958. %doi:10.1093/mnras/staa1325

    \bibitem[Weilbacher et al.(2014)]{2014ASPC..485..451W} Weilbacher, P.~M., Streicher, O., Urrutia, T., et al.\ 2014, Astronomical Data Analysis Software and Systems XXIII, 485, 451. %doi:10.48550/arXiv.1507.00034
    
    \bibitem[Weilbacher et al.(2020)]{2020A&A...641A..28W} Weilbacher, P.~M., Palsa, R., Streicher, O., et al.\ 2020, \aap, 641, A28. %doi:10.1051/0004-6361/202037855

    \bibitem[Wellons et al.(2015)]{2015MNRAS.449..361W} Wellons, S., Torrey, P., Ma, C.-P., et al.\ 2015, \mnras, 449, 361. %doi:10.1093/mnras/stv303

    \bibitem[Whitaker et al.(2012)]{2012ApJ...754L..29W} Whitaker, K.~E., van Dokkum, P.~G., Brammer, G., et al.\ 2012, \apjl, 754, L29. %doi:10.1088/2041-8205/754/2/L29

    \bibitem[Whitaker et al.(2017)]{2017ApJ...838...19W} Whitaker, K.~E., Bezanson, R., van Dokkum, P.~G., et al.\ 2017, \apj, 838, 19. %doi:10.3847/1538-4357/aa6258

    \bibitem[Cleveland \& Devlin(1988)]{Cleveland1988LocallyWR} William S. Cleveland and Susan J. Devlin.\ 1988, Journal of the American Statistical Association, 83, 596-610 %I made this reference

    \bibitem[Wisnioski et al.(2015)]{2015ApJ...799..209W} Wisnioski, E., F{\"o}rster Schreiber, N.~M., Wuyts, S., et al.\ 2015, \apj, 799, 209. %doi:10.1088/0004-637X/799/2/209

    \bibitem[Wuyts et al.(2011)]{2011ApJ...742...96W} Wuyts, S., F{\"o}rster Schreiber, N.~M., van der Wel, A., et al.\ 2011, \apj, 742, 96. %doi:10.1088/0004-637X/742/2/96



\end{thebibliography}
\end{document}